\documentclass[fleqn,usenatbib]{mnras}

\usepackage{newtxtext,newtxmath}

\usepackage[T1]{fontenc}
\usepackage{ae,aecompl}

\usepackage{graphicx}	
\usepackage{amsmath}	
\usepackage{amssymb}	

\usepackage{lscape}     

\title[Metal absorption systems near reionization]{Modelling intergalactic low ionisation metal absorption line systems near the epoch of reionization}

\author[T. Suarez \& A. Meiksin]{Teresita Suarez $^{1}$ \thanks{Contact e-mail: \href{mailto:suarez@roe.ac.uk}{tsuarez@ed.ac.uk}}
\& Avery Meiksin $^{1}$
\\
$^{1}$SUPA\thanks{Scottish Universities Physics Alliance}, The Royal Observatory, Edinburgh, Blackford Hill, Edinburgh EH9 3HJ, UK
}

\date{Accepted XXX. Received YYY; in original form ZZZ}

\pubyear{2020}

\begin{document}
\label{firstpage}
\pagerange{\pageref{firstpage}--\pageref{lastpage}}
\maketitle

\begin{abstract}
We interpret observations of intergalactic low ionisation metal absorption systems at redshifts z $\gtrsim$5 in terms of pressure-confined clouds. We find clouds confined by the expected pressure of galactic haloes with masses $11<\log M_h/h^{-1}M_\odot<12$ provide a good description of the column density ratios between low ionisation metal absorbers. Some of the ratios, however, require extending conventional radiative transfer models of irradiated slabs to spherical (or cylindrical) clouds to allow for lines of sight passing outside the cores of the clouds. Moderate depletion of silicon onto dust grains is also indicated in some systems. The chemical abundances inferred span the range between solar and massive-star dominated stellar populations as may arise in starburst galaxies. The typical H{\sc i} column densities matching the data correspond to Damped Lyman-$\alpha$ Absorbers (DLAs) or sub-DLAs, with sizes of 40~pc to 3~kpc, gas masses $3.5<\log M_c/M_\odot<8$ and metallicites $0.001-0.01Z_\odot$. Such systems continue to pose a challenge for galaxy-scale numerical simulations to reproduce.

\end{abstract}




\section{Introduction}

Cosmic reionization remains one of the paramount unsolved problems of modern cosmology. Observations of high redshift Quasi-Stellar Objects (QSOs) show the intergalactic medium (IGM) was reionized by $z=5$ \citep{2015MNRAS.447.3402B, 2018MNRAS.479.1055B}. At higher redshift, evidence from QSO \citep{Davies2018, Greig2017} and Lyman Break Galaxy \citep{2018ApJ...856....2M} spectra suggest the IGM was largely neutral at $z\gtrsim7.5$. Measurements of the Cosmic Microwave Background (CMB) require the IGM to have been ionised no earlier than $z\simeq8.4$, and suggest a midpoint reionization epoch of $z_{\rm re}=7.7\pm0.7$ \citep{2018arXiv180706209P}.

The sources responsible for reionizing the IGM are unknown. Widely held to be star-forming galaxies, current evidence shows that the ionising emissivity of observed sources at $z>5$ still falls short of what is required \citep{Bouwens2015, Finkelstein2019ApJ...879...36F}. Whilst extending the luminosity function of observed galaxies to fainter systems produces the emissivity required to reionize the Universe, it is not clear that a sufficient fraction of photons manage to escape the galaxies \citep{2019MNRAS.485...47P}. Some studies suggest that galaxies hosting active galactic nuclei contribute significantly to the ultraviolet background near the end of the reionization epoch at $z<7$ \citep{Giallongo2015, Finkelstein2019ApJ...879...36F},  assuming essentially full escape of the ionising radiation produced.

Although the full complement of galaxies responsible for reionizing may be too dim to detect, they may be indirectly discovered through their impact on the IGM and their nature inferred. Models suggest material lost from the galaxies in winds enrich the IGM in metals \citep[for reviews, see][] {McQuinnreview, 2017ARA&A..55..389T}. The chemical abundances of the inter-stellar and circum-galactic medium of the galaxies are expected to reflect the abundances of the stellar populations of the galaxies. The metal abundances of this enriched gas thus provide valuable insight into the nature of the earliest galaxies.

Numerical simulations suggest the first generation of stars after the Big Bang were very massive and were composed of pristine hydrogen and helium gas (Population~III stars). Subsequent generations of stars formed from chemically enriched material but they are presumed to be metal-poor type stars  (Population II stars), and their chemical abundance patterns will depend on their environments \citep[eg][]{Starkenburg10.1093/mnras/stw2873}.
Modelling the stellar populations is further complicated by anomalies in the abundance patterns of metal-poor stars \citep{2002Natur.419..904C, Frebel2005Natur.434..871F, Norris2007ApJ...670..774N}, which show large overabundances in carbon and nitrogen of unclear origin. One suggestion is that gaseous regions with such overabundances are preferentially selected for forming stars as they provide the most efficient conditions for gas to cool and fragment and form stars \citep{2007MNRAS.380L..40F}.

Intergalactic metal absorption systems in the spectra of background quasars potentially provide a powerful probe of the composition of early stars and their environments. Measurements of element abundance ratios have been used to infer the nature of stellar populations in young galaxies \citep[eg][]{Becker2011-12}.

At high redshifts the forest is too highly absorbed, becoming almost completely opaque by $z\sim 6$. The saturation of the absorption lines makes it impossible to make direct estimates of the metallicities for these systems. Instead, ionisation models of the absorption systems are necessary.

Early galaxies are expected to be surrounded by a reservoir of gas, from which they accrete and form stars. High H{\sc i} column density intervening absorbers, especially sub-Damped Lyman Alpha (sub-DLAs, having $10^{19}<N$(H{\sc i})$<10^{20.3}\,{\rm cm}^{-2}$) and Damped Lyman Alpha (DLAs, $N$(H{\sc i})$>10^{20.3}\,{\rm cm}^{-2}$), may serve as probes of this gas.
DLAs are found to have metallicties typically below 0.1 Z$_\odot$ at intermediate redshifts $z\sim2-4$ \citep{1997ApJ...486..665P, Prochaska2003ApJS..147..227P} and rising above $0.1$ solar metallicities at low redshift $z<1.5$ \citep{2005AJ....129....9R}. These previous studies indicate that the metallicities of DLAs decrease with increasing redshift \citep{Wolfe2005ARA&A..43..861W}. However, observations show a 'floor' in the metallicity of about $0.002Z_\odot$ out to $z\sim 5$ \citep{Wolfe2005ARA&A..43..861W, Rafelski2012ApJ...755...89R}, with rare exceptions \citep{2010MNRAS.406.1435E}. Many metal-poor DLAs at intermediate redshifts $z<4$ exhibit $\alpha$-enhanced metallicities \citep{CookeRJ2015, Rafelski2012ApJ...755...89R}.

Some systematic differences are found. Unlike their lower redshift counterparts, sub-DLA and DLA systems at $z>5$ lack strong absorption by high ionisation species like CIV and SiIV, consistent with an evolving metallicity \citep{Becker:2011}. On the other hand, low ionisation species like C{\sc ii}, Si{\sc ii}, Fe{\sc ii} and O{\sc i} show little evolution in their column density ratios over $2<z<6$ in some samples, with values similar to low metallicity non-carbon enhanced Pop~II stars \citep{Becker2011-12}. Other samples, however, do show evidence for mild metallicity evolution for undepleted elements, with some systems at $z\sim5$ having anomalously low [C/O] abundances \citep{Poudel2018}.

The low metallicities may be explained in a scenario in which star-forming galaxies enrich their environments over time. Galaxies build up their stellar content by recycling mass from gas expelled by evolved stellar populations (which are metal-rich), but also from gas accreted from the IGM. The accretion is believed regulated by feedback processes that may expel material from galaxies, limiting star formation. Intervening metal absorption systems may thus shed insight on the processes that enrich the stellar populations of galaxies as well as pollute the environments of galaxies, such as by supernovae-driven galactic winds \citep[see][for a review]{2015ARA&A..53...51S}.

Feedback may regulate not only the metallicity of the gas and stars, but their abundances as well. Abundances dominated by core-collapse supernovae will be $\alpha-$enhanced compared with solar abundances, an effect that may be enhanced in more massive galaxies as a result of feedback by Active Galactic Nuclei (AGN) \citep{Segers2016}.

Cool clouds in the Circumgalactic medium (CGM) may be highly transient, requiring constant replenishment of new material from their host galaxies. Models for the origin of the clouds include formation in gaseous outflows, inflows and \textit{in-situ} formation \citep[eg][]{2012MNRAS.420..829O, 2015MNRAS.448..895S, Suarez2016MNRAS.462..994S, 2017MNRAS.471..690T}. The physics of the cool CGM is unresolved in cosmological hydrodynamic simulations. The physical scale of the clouds is estimated from observations to be as small as tens of parsecs, and possibly smaller, at low redshifts $z<3$ \citep{Prochaska2009ApJ...690.1558P, Lan2017ApJ...850..156L}. Such short scales pose a challenge for galaxy-scale hydrodynamical simulations \citep{Peeples2019, Sparre2019MNRAS.482.5401S}. Recourse must instead be made to semi-analytic modelling. Estimated cloud masses for spherical models of circumgalactic sub-DLA and DLA systems pressure-confined in galactic haloes range from 100 to 10$^8$ M$_\odot$, increasing with H{\sc i} column density, and with sizes from 10 to several hundred parsecs \citep{LanMo2019}.

In this paper, we investigate the metal abundances of high redshift ($z>5$) low ionisation absorption systems in the context of pressure-confined clouds with the view of learning the nature of the stellar populations arising during cosmic dawn, as reflected by the surrounding gas they enriched. All of the systems we model are likely sub-DLA or DLAs, although confirming direct H{\sc i} absorption measurements are not possible for many. Another advantage of examining systems at $z>5$ is their reduced metallicities \citep{2016ApJ...830..158M}, mitigating the complication of dust depletion and permitting a cleaner test of the pressure-confined cloud model.

We use the radiation transfer code {\sc cloudy} \citep{Ferland1998, Ferland2017} to model the column density ratios. Rather than following the general practice of using the results from slab models as is, in this paper we instead roll the clouds into cylinders to approximate the changing relative ionisation fractions between elements expected in spherical systems. This extends the dynamic range of predicted column density ratios by considering lines of sight passing through a range of projected separations from the cloud centres, and may explain some of the scatter in the measured ratios. We also consider both solar chemical abundance ratios and abundances more typical of Pop~II dominated starburst galaxies \citep{HamannFerland:1993}.

This paper is organised as follows. In the next section, we describe the configuration of the models and the parameterisation we use. We then discuss the characteristics and motivation of modelling absorption systems as pressure-confined clouds in Sec. \ref{results}. We compare our models to current available observations of metal absorption systems at $z\gtrsim5$ in Sec. \ref{observations}. We discuss our results in Sec. \ref{discussion} and present our conclusions in Sec. \ref{conclusions}.

\begin{figure}
    \centering
    \includegraphics[trim=10cm 0cm 10cm 0cm, clip, width=\columnwidth, height=14 cm]{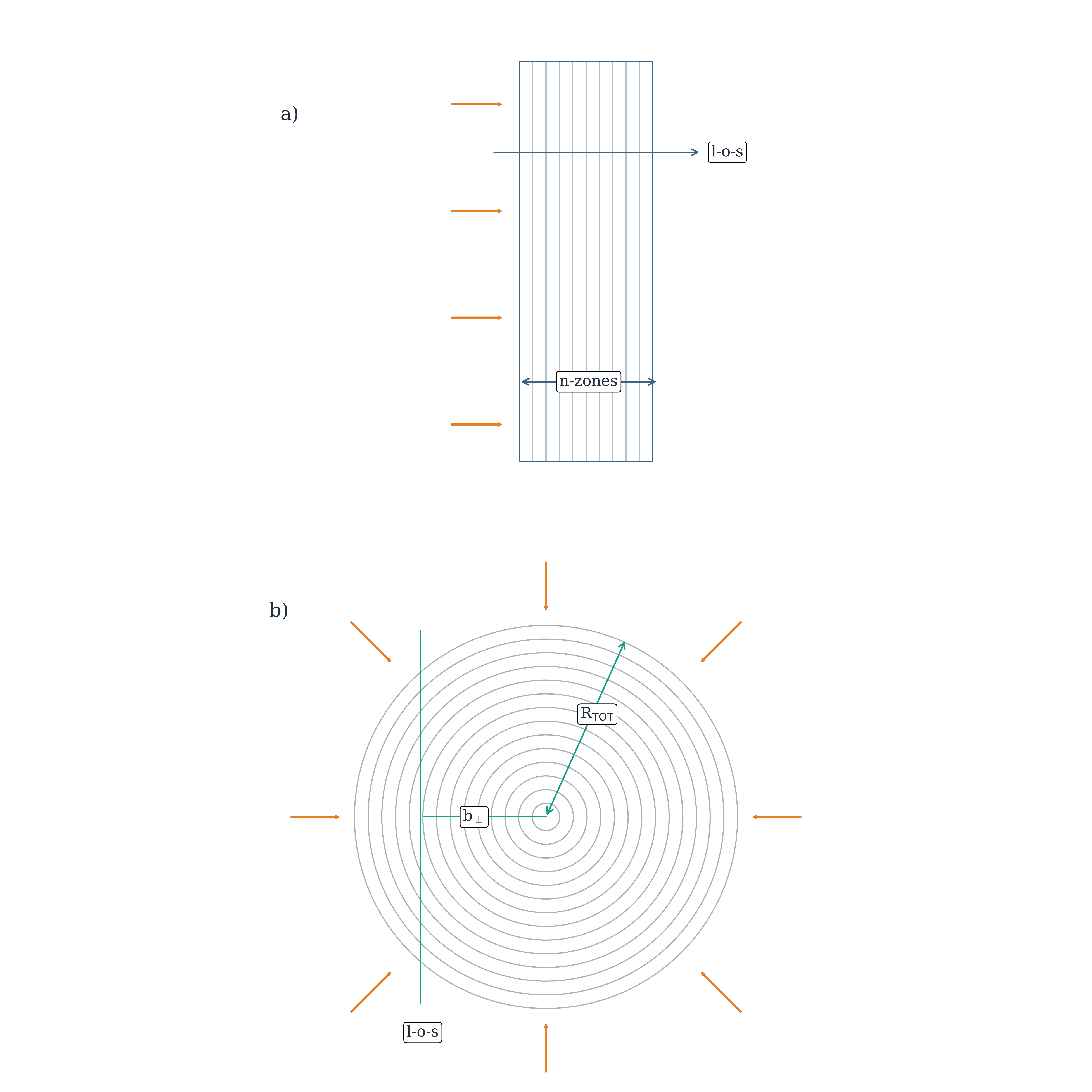}
    \caption{a) Slab geometry. This model assumes an infinitely broad flat cloud with an external radiation field impinging perpendicularly on one side. b) Circular cloud: radiation hits the cloud from all directions perpendicular to its surface. }
    \label{fig:SlabCircle}
\end{figure}

\section{Configuration of the models}
\label{Methods}

\begin{table*}
	\centering
	\begin{tabular}{lccccr} 
		\hline
		log $n_\mathrm{H}$ & log $N$(H{\sc i}) & Radiation& Redshift & Metallicity & Geometry\\
		cm$^{3}$ & cm$^{-2}$ &  background & $z$ & $Z$ & \\		
		\hline
		[-3.0,-1] & 20 & H\&M 2012 & 5 &solar & slab\\
		          & 21 & H\&M 2005 & 6 &starburst & circular\\
		  &   &    &   & &\\
		\hline
	\end{tabular}
    		\caption{{\sc cloudy} input parameters to model the physical properties of isobaric clouds in the CGM; $n_\mathrm{H}$, $N$(H{\sc i}) and H\&M are the gas hydrogen density, the neutral hydrogen density and the radiation continuum by \citet[][2005]{HM2012}, respectively. (Each column represents options independent of the option chosen in any other column.)
	}
	\label{tab:parameters}
\end{table*}

\begin{table}
    \centering

\begin{tabular}{rl|rl}
\hline
 \hline
 \multicolumn{2}{|c|}{ solar ($\odot$)  }  &    \multicolumn{2}{|c|}{starburst (Sb)   }    \\
\hline
   Element  &   log &    Element  &   log \\
\hline
H 	&	  0.0000 	&	H 	&  0.0000 	\\
C 	&	 -6.611 	&	C 	& -7.104	\\
Si	&	 -7.460 	&	Si	& -7.465	\\
O 	&	 -6.310 	&	O 	& -6.182	\\
Fe	&	 -7.550 	&	Fe	& -8.234	\\
Al	&	-8.530 	&	Al	& $\dots$	\\
Mg	&	-7.460 	&	Mg  & -7.466	\\
\hline
\hline
\end{tabular}
    \caption{Abundances specified by {\sc cloudy} for the composition of default solar composition abundances and for a stellar population dominated by massive stars (using the \textit{starburst abundances} option in {\sc cloudy}). The \textit{starburst abundances} values are for an evolving starburst galaxy from \citet{HamannFerland:1993}. The metallicity of the gas is relative to a solar Si abundance, with metallicity set at $Z=0.001Z_\odot$.
}
    \label{tab:abundances}
\end{table}

We model clouds in the circumgalactic medium of galaxies using the spectral synthesis code {\sc cloudy} v.17.01 \citep{Ferland1998, Ferland2017}.

We work under the assumption that non-isobaric clouds would restore pressure balance by fragmenting into smaller clumps of gas \citep{LanMo2019}. The models described here adopt the pressure at the illuminated face of the cloud and force the pressure to be kept constant throughout the cloud. We refer to the systems described here as pressure-confined models.

Modelling metal absorption systems using {\sc cloudy} allows specification of the radiation field shape and intensity. In our models we assume a UV background with contributions from both quasars (QSOs) and galaxies using a background continuum from a Haardt \& Madau table \citep{HM2012} at different redshifts. We initially explored a wide range of cloud densities values, $n_\mathrm{H}$, then narrowed the range by requiring physical stability and to match measured metal column density ratios and limits. As we shall see, the cloud densities are consequently restricted to the range $-3 < \log n_\mathrm{H} < -1 $. This is the density at the surface of the cloud, as it will usually increase interior to the cloud to maintain constant pressure.

These models stop the calculation when reaching a column density of neutral hydrogen equivalent to $10^{20}-10^{21}\,{\rm cm}^{-2}$ from surface to centre, to approximate sub-DLA and DLA systems. We reference the model with the H{\sc i} column densities set to one of these values in tables and figure legends. In agreement with previous modelling, we find that the low ionisation metal column density ratios require high neutral hydrogen column densities typical of DLAs \citep[eg][]{Becker:2011}. We also explore the effect of metal abundances by examining both models with solar abundances and with chemical abundances typical of an evolving starburst in a massive elliptical galaxy with a top-heavy stellar initial mass function, based on the galactic chemical evolution model M5a of \citet{HamannFerland:1993}, implemented using the \lq Abundances starburst\rq\ option in {\sc cloudy}.

The geometric configuration used by {\sc cloudy} is a spherical gaseous shell surrounding a central source. A plane parallel slab with a radiation field impinging perpendicularly to the surface is approximated by asserting an extremely large radius for the shell compared with its thickness (the slab geometry). The inverse problem of a spherical cloud embedded in an external radiation field, the case of interest to the study of optically thick intergalactic absorption systems, is not available. Models of intergalactic absorption systems are usually computed using the slab geometry setting, and column densities are computed along a normal to the slab surface. This fails to capture the actual ionisation layering within a spherical cloud and the column density ratios that would arise along lines of sight with impact parameters displaced from the cloud centre. To approximate such lines of sight, we \lq\lq roll\rq\rq\ the slab into a cylinder, allowing us to compute the column densities for ions along lines of sight displaced from the cloud centre, as shown in Fig.~\ref{fig:SlabCircle}. As we shall show, variations in ion column density ratios arise as the impact parameter is changed. The variations will also depend on the confining pressure and size of the absorbers, so that measured column density ratios may be used to constrain these properties.

In Table~\ref{tab:parameters} we summarise the values of the model parameters used. They are restricted by several physical considerations, as described in the following section.

\section{Physical characteristics of pressure-confined clouds}
\label{results}

In this section we analyse the physical characteristics of pressure-confined clouds. We describe the variations in ionisation fractions with radius based on {\sc cloudy} models, and describe some of the considerations limiting the cloud properties.

\begin{figure}
    \centering
    \includegraphics[width=\columnwidth]{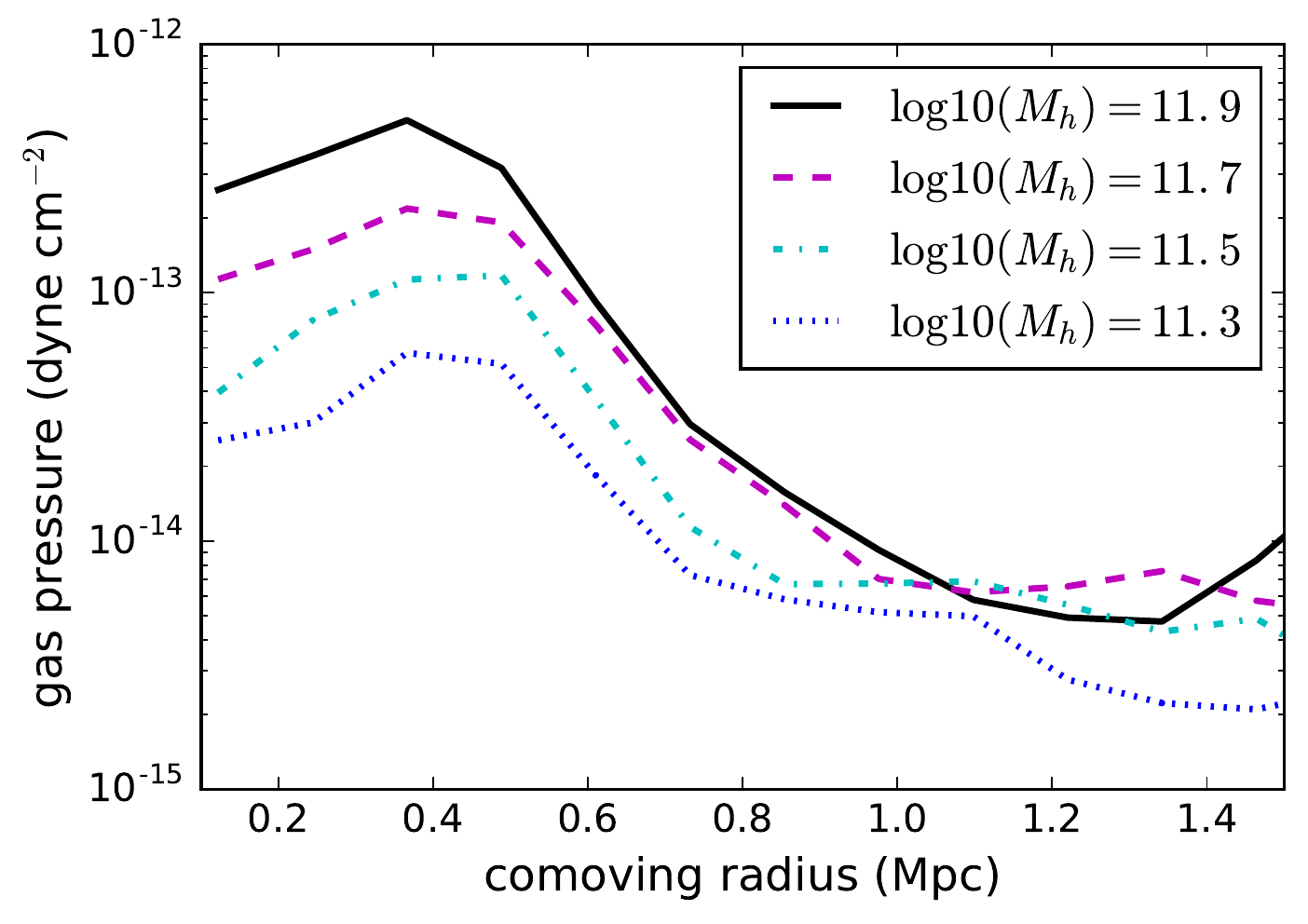}
    \caption{Gas pressure around dark matter haloes of masses between $10^{11}$ and $10^{12}\,h^{-1}M_\odot$ at $z\simeq6$, from the Sherwood simulation suite \citep{2017MNRAS.464..897B}.
    }
    \label{fig:phaloes}
\end{figure}

Because galaxies are expected to reside in extended, overdense structures, the associated gas will be at much higher pressure than in the diffuse IGM. Although there are no direct measurements of the gas pressure in the haloes of high redshift star-forming galaxies, estimates may be made from cosmological simulations. Radially averaged gas pressure profiles at $z\simeq6$ from the Sherwood simulation suite \citep{2017MNRAS.464..897B} are shown in Fig.~\ref{fig:phaloes} for a $\Lambda$CDM cosmology. Galactic wind feedback is included, following the prescription of \citet{2013MNRAS.428.2966P}, which is found to reproduce the rising HI absorption signature measured in the CGM of star-forming galaxies at $z\simeq3$ \citep{2017MNRAS.468.1893M}. Thermal gas pressures at distances $20-100\sim$kpc (proper) from the halo centres of $10^{-14}-10^{-12}\,{\rm erg}\,{\rm cm^{-3}}$ are predicted for systems with halo masses between $10^{11}-10^{12}\,h^{-1}\,M_\odot$, the characteristic halo mass range believed associated with DLAs for at least moderate redshift systems \citep{2012JCAP...11..059F, 2014MNRAS.440.2313B}.

\begin{figure}
    \centering
    \includegraphics[width=\linewidth]{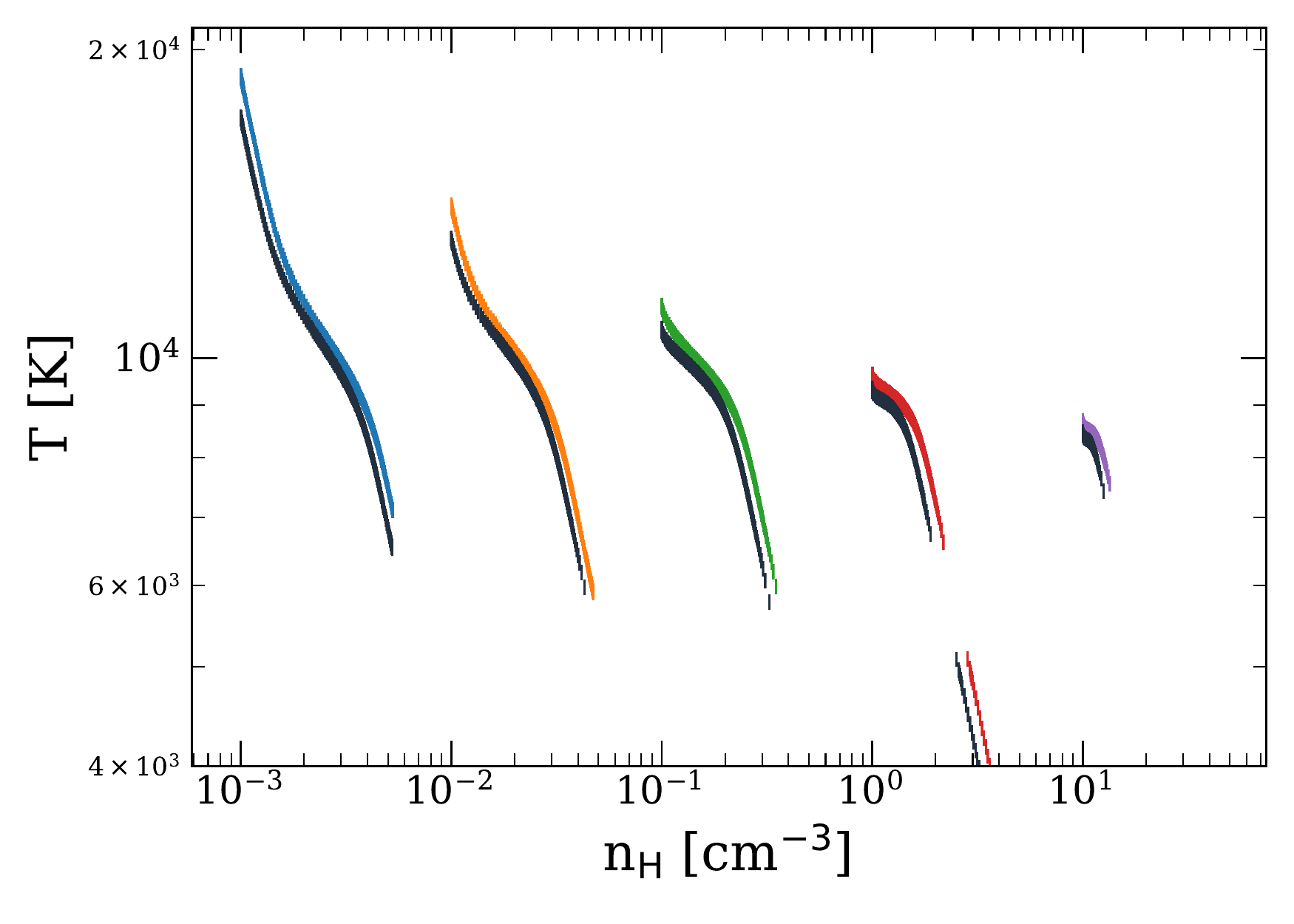}
    \includegraphics[width=\linewidth]{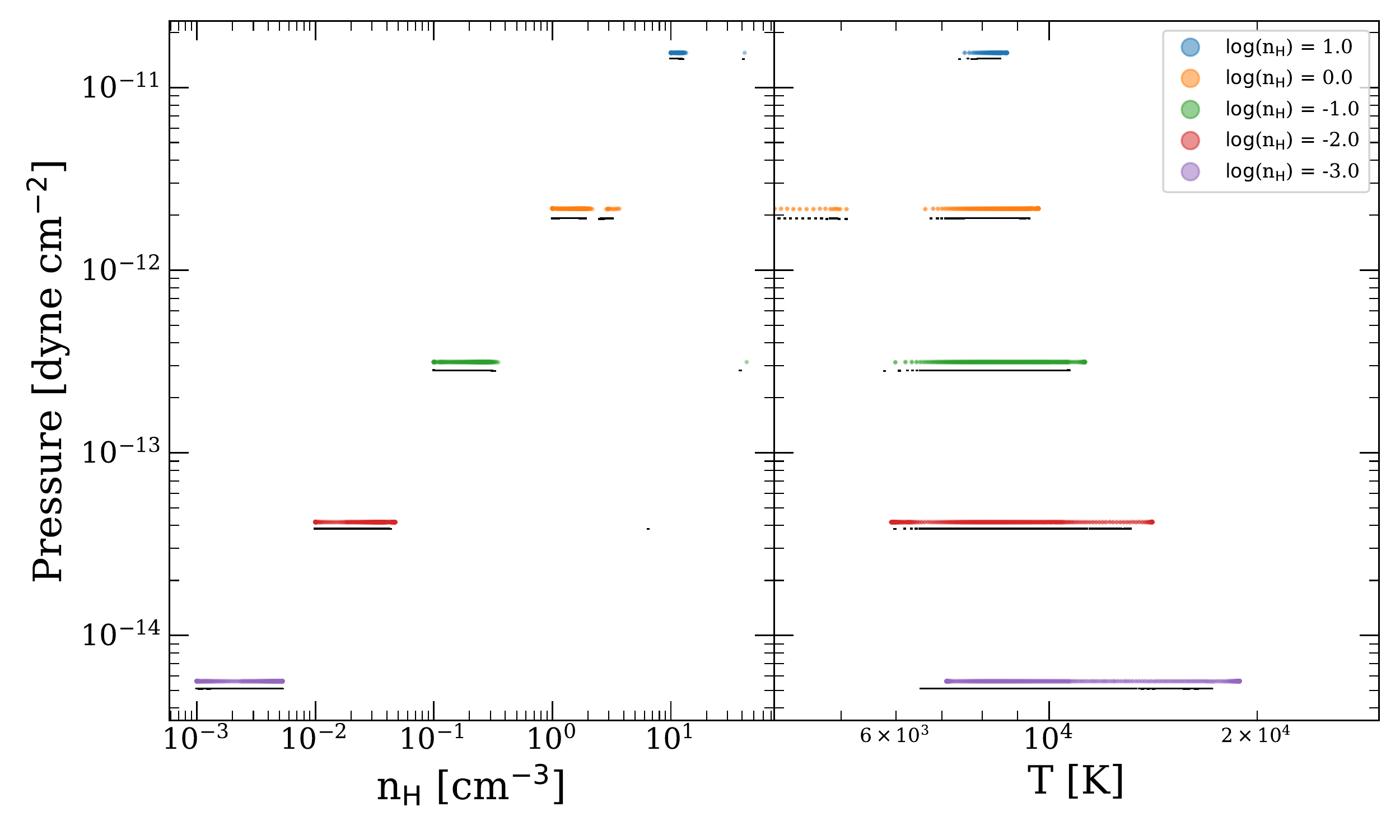}
    \caption{Physical properties of pressure-confined clouds. The top panel shows the temperature as a function of hydrogen density internal to the clouds. The bottom panel shows the ranges in density and temperature for selected isobaric models. The coloured lines are for clouds photoionised by a metagalactic field at $z=6$ \citep{HM2012}. Black lines are models using the metagalactic field at $z=7$.
    }
    \label{fig:pressure_temp_dens}
\end{figure}

\begin{figure}
    \centering
    \includegraphics[trim=1cm 1.8cm 2.5cm 3cm,clip,width=\columnwidth]{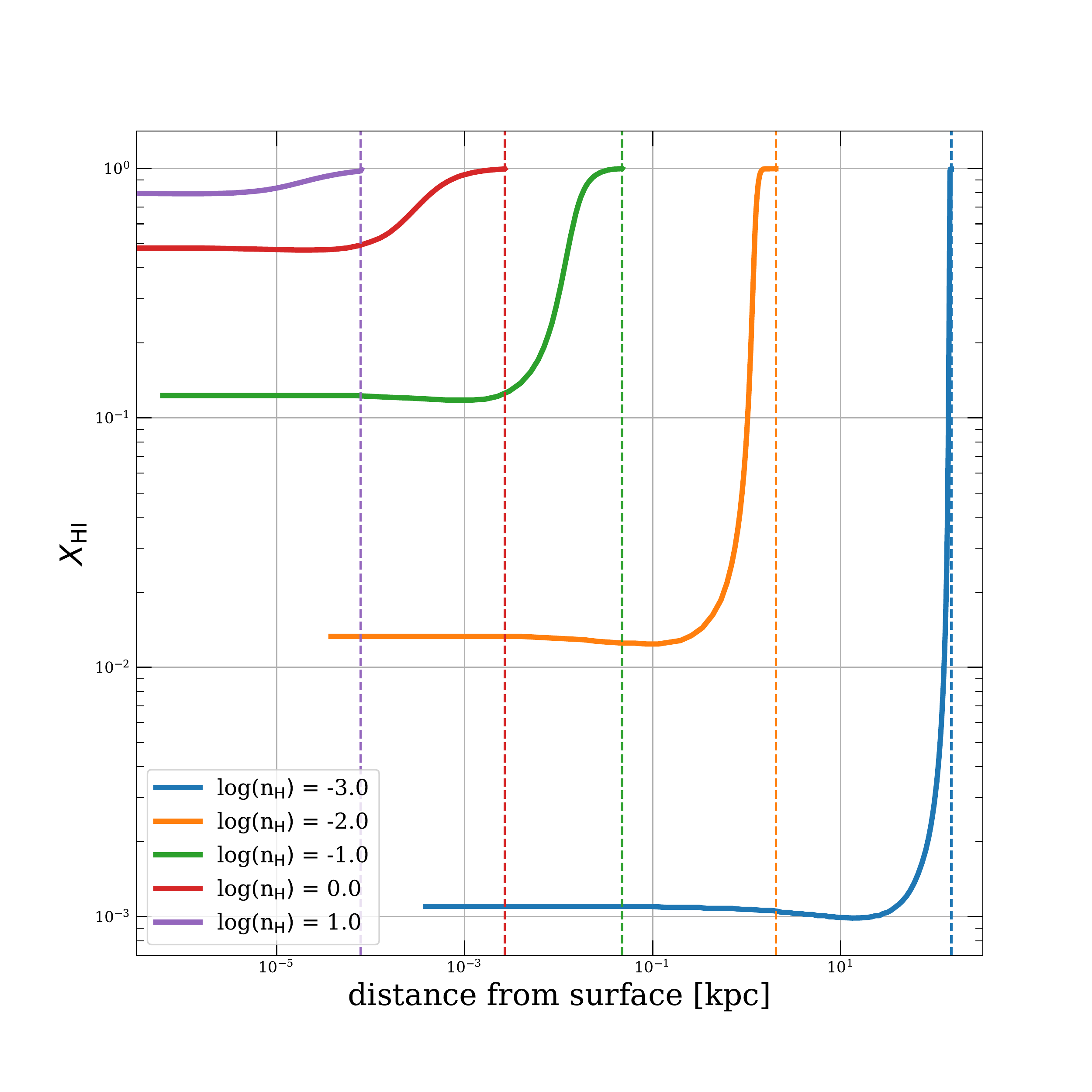}
    \caption{Fraction of neutral hydrogen (H{\sc i}) as a function of distance from the cloud surface. The line colours correspond to different values in the surface hydrogen density. Lower hydrogen densities produce larger clouds. The hydrogen densities $n_\mathrm{H}$ are in cm$^{-3}$.
    }
    \label{fig:HI_fraction}
\end{figure}

\begin{figure}
    \centering
    \includegraphics[width=\columnwidth]{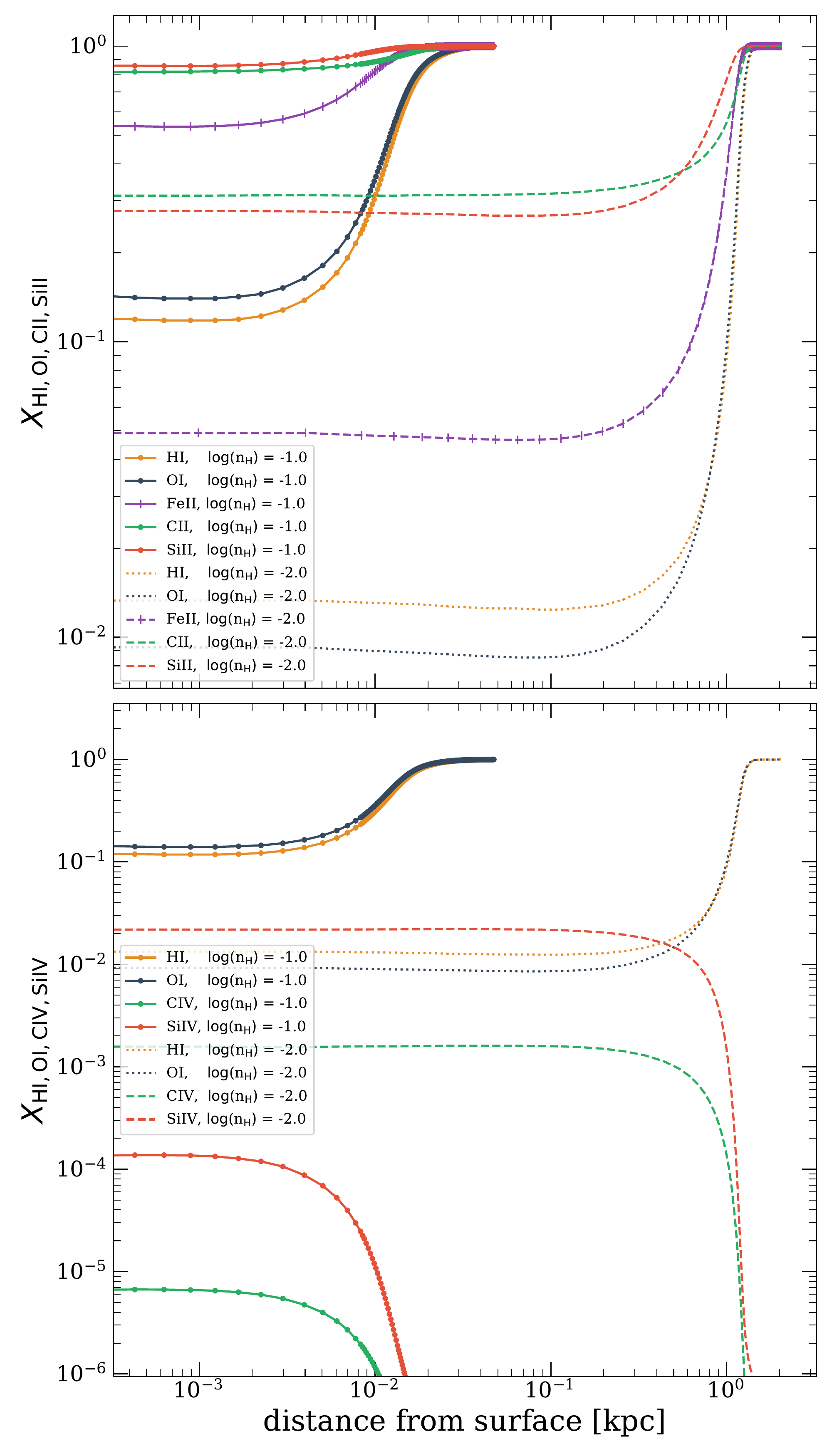}
\caption{An example of the ionisation fractions of elements, $X_\mathrm{ion}$, as a function of distance from the cloud surface. The top panel shows the ionisation fraction of H{\sc I} and the low-ionisation elements Fe{\sc ii}, C{\sc ii}, O{\sc i} and Si{\sc ii} for two different hydrogen densities $n_\mathrm{H}$ = 0.1, 0.01~cm$^{-3}$. The bottom panel shows the high-ionisation ions C{\sc iv} and Si{\sc iv}. We show the fractions of H{\sc i} and O{\sc i} for comparison. We set the densities also to $n_\mathrm{H}$ = 0.1, 0.01~cm$^{-3}$. These fractions were generated by a model with $N$(H{\sc i})=$10^{20}$~cm$^{-2}$, metallicity from massive stars and a radiation field at $z=6$ using a H\&M 2012 model.}
    \label{fig:metIonStage}
\end{figure}

Following earlier models \citep{LanMo2019, MoMiralda1996ApJ...469..589M}, we assume the clouds are confined by the pressure of the haloes. By choosing the hydrogen density at the surface of the cloud, a pressure is computed at the illuminated cloud surface required for thermal equilibrium. The pressure is forced to remain constant at this value throughout the cloud interior.

Fig.~\ref{fig:pressure_temp_dens} illustrates the behaviour of the density and temperature interior to the clouds for different assumed densities at the surface, along with the required pressure resulting from thermal equilibrium. The top panel of Fig.~\ref{fig:pressure_temp_dens} shows the temperature as a function of internal hydrogen density, $n_\mathrm{H}$ for different clouds. The bottom panel of Fig.~\ref{fig:pressure_temp_dens} shows the ranges in hydrogen density and temperature interior to each cloud model for models with different pressures. The coloured curves are for models photoionised by the metagalactic radiation field of \citet{HM2012} at $z=6$, and the black lines shows how these curves would change with variations in the ultra-violet background (UVB), for which we take the extragalactic background from \citet{HM2012} at redshift $z=7$ as an illustration. Only mild sensitivity to the external radiation field is found.

Fig.~\ref{fig:HI_fraction} depicts the variation in the neutral hydrogen fraction with distance from the cloud surface for different values of the surface hydrogen density $n_\mathrm{H}$. The higher $n_\mathrm{H}$ is, the smaller the cloud radius.
The difference in radii spans six orders of magnitude.
A brief description of these models is:\ the cloud with surface density $n_\mathrm{H}$ = 10~cm$^{-3}$ (purple curve) has a neutral hydrogen fraction of $X_\mathrm{HI} = 0.8$ at the surface.
The cloud radius is $<1$~pc. Next, a surface density $n_\mathrm{H}$ = 1~cm$^{-3}$ gives a cloud with radius $R_c \sim$2.5~pc. It is about half ionised at its surface, quickly rising to being almost fully ionised at half its radius. Clouds with $n_\mathrm{H} = 0.1\,{\rm cm}^{-3}$ (green curve) reach a size of $\sim50$~pc. The surface neutral fraction is low, $X_\mathrm{HI}\simeq0.1$, reaching 0.9 at the half radius. Models with $n_\mathrm{H} = 0.01\,{\rm cm}^{-3}$ (yellow curve) show a more bimodal behaviour. The cloud remains mostly neutral from its centre to its half radius. Outside the half radius, the neutral fraction climbs rapidly, reaching $X_\mathrm{HI}$ = 1 at three quarters of the cloud radius. Finally, the model with $n_\mathrm{H}$ = 0.001~cm$^{-3}$ (blue curve) shows a size of $R_c =$ 150~kpc. The fraction of neutral hydrogen $X_\mathrm{HI}$ remains close to zero almost throughout the cloud, transitioning rapidly to $X_\mathrm{HI}$ =1 only within 10~kpc of its centre.

Fig.~\ref{fig:metIonStage} shows the variation in ionisation fraction with distance from the cloud surface for low-ionisation elements (O{\sc i}, Fe{\sc ii}, C{\sc ii} and Si{\sc ii}; top panel) and high-ionisation elements (C{\sc iv}  and Si{\sc iv}; bottom panel). For comparison we also show the ionisation fraction of H{\sc i}. The values shown in Fig.~\ref{fig:metIonStage} are for a cloud with a column density of log $N$(H{\sc i}) = $10^{20}\,{\rm cm}^{-2}$, starburst abundances and metallicity of 0.001 relative to solar. The \citet{HM2012} radiation field at $z=6$ is used.

The trends in the ionisation fractions depend on the cloud model. For the higher density case, $n_\mathrm{H}=0.1\,{\rm cm}^{-3}$, the cloud radius is  $R_c\sim 50$~pc. The ionisation fractions of C{\sc ii} and Si{\sc ii} are high, remaining at $X_\mathrm{CII,SiII}>0.8$, through the entire cloud. The ratio of $N$(Si{\sc ii}) to $N$(C{\sc ii}) varies with depth, with $X_\mathrm{SiII}\sim 1.0$ one-fifth the way into the cloud, while $X_\mathrm{CII}\sim1.0$ only within the inner half of the cloud. The fractions of H{\sc i} and O{\sc i} start to grow $\sim10$~pc from the surface. Having nearly identical ionisation potentials, they closely track each other, reaching $X_\mathrm{HI,OI}>0.8$ at the half radius. The high-ionisation elements (C{\sc iv} and Si{\sc iv}),  bottom panel of Fig.~\ref{fig:metIonStage}, are negligible, both below $2\times10^{-4}$, and vanishingly small in the inner half of the cloud.

Somewhat different trends for low-ionisation ions are found for a surface hydrogen density $n_\mathrm{H} = 0.01\,{\rm cm}^{-3}$. This cloud is substantially larger, with a radius of about $\sim2$~kpc. Although the C{\sc ii} fraction slightly exceeds the Si{\sc ii} fraction near the cloud surface, the Si{\sc ii} fraction increasingly dominates one-third the way into the cloud until they both reach unity at the cloud centre. The H{\sc i} and O{\sc i} fractions are at the percent level from the surface to half radius, both increasing rapidly thereafter. Compared with the previous denser cloud, the ionisation fractions of the high-ionisation elements C{\sc iv} and Si{\sc iv} are much higher, but plummet to vanishingly low levels within the inner half of the cloud (bottom panel of Fig.~\ref{fig:metIonStage}).

We also test the stability of the cloud by calculating the Jeans length ($\lambda _\mathrm{J}$) for each model,
\begin{equation}
     \lambda_\mathrm{J} \simeq c_{\mathrm{s}}\; (4\pi G m_{\mathrm{p}} n_{\mathrm{H}})^{-1/2}
\end{equation}
 \noindent where $c_{\mathrm{s}}$ is the speed of sound, $G$  is the gravitational constant and $m_{\mathrm{p}}$ is the mass of a proton \citep{Spitzer1978ppim.book.....S}. The values for the radii and masses of the clouds and Jeans length for relevant models are summarised in the model tables below. Only Jeans stable systems are used. (Note that \lq log\rq\ values are in base 10.) A lower limit on cloud mass imposed by thermal heat conduction is estimated at around $10^3-10^4\,M_\odot$, below which pressure-confined clouds evaporate on the timescale of tens to several hundred million years \citep{2017MNRAS.470..114A, LanMo2019}.


\begin{figure*}
    \centering
    \includegraphics[trim=2cm 1.5cm 4.5cm 3cm,clip,width=\linewidth]{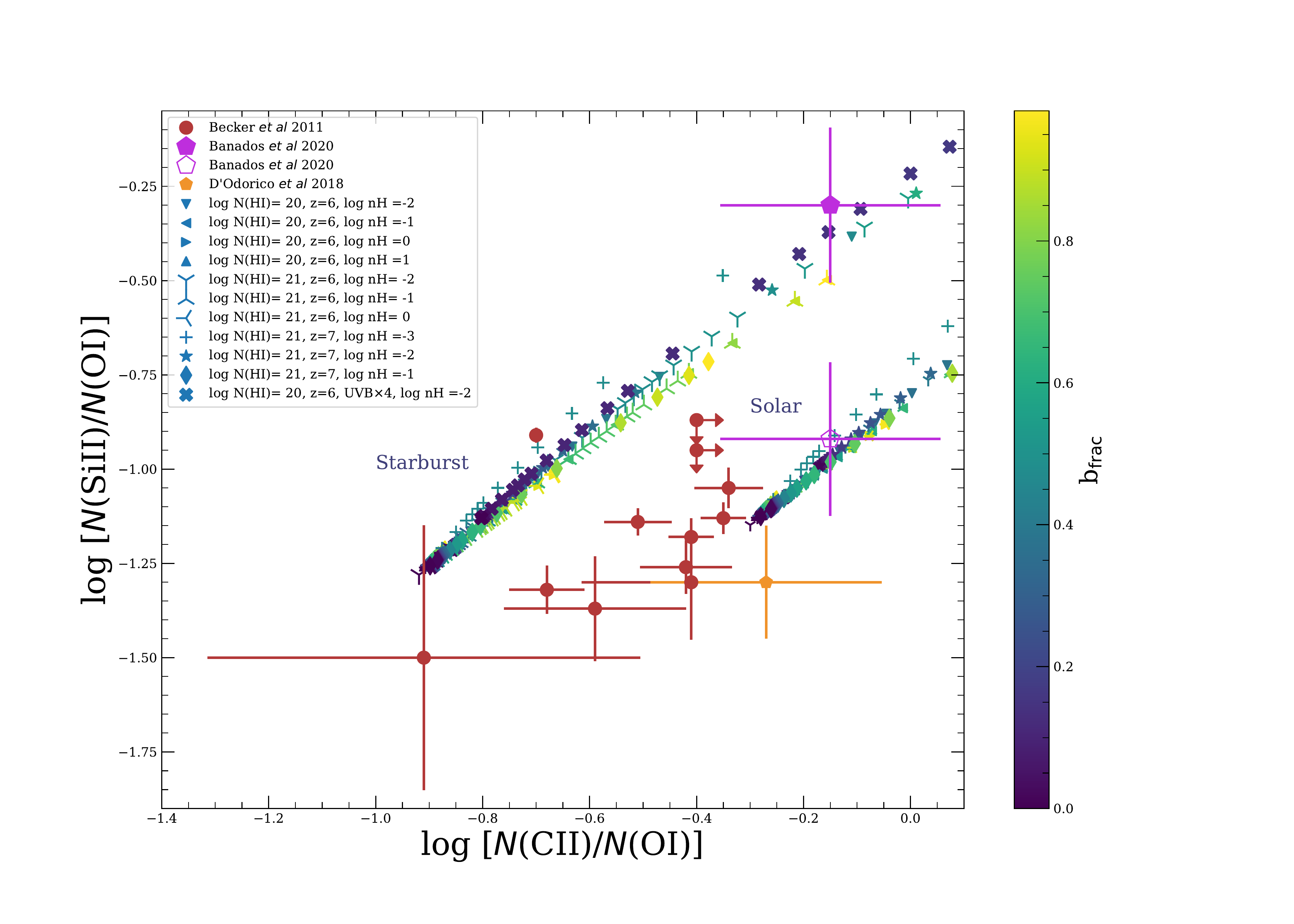}
\caption{Column density ratios between Si{\sc ii}, C{\sc ii} and O{\sc i} for data from \citet{Becker2011-12} (red circles), \citet{Banados2019ApJ...885...59B} (full and empty pink pentagons; the Si{\sc ii} values are upper limits, see text) and  \citet{D_Odorico_2018} (orange pentagon). Markers in colour are {\sc cloudy} models corresponding to the legend, with a clear division between predictions for different abundances. Models have slab neutral hydrogen column densities $N$(HI)=20 cm$^{-2}$ and $N$(H{\sc i})=21 cm$^{-2}$. The surface hydrogen density ($n_\mathrm{H}$) varies between 0.001 and 1 cm$^{-3}$. All models use a radiation background from \citet{HM2012} at $z=6$ and 7. The colour bar indicates the impact parameter of the line-of-sight measurement through the cloud, with dark blue corresponding to small impact parameters and light yellow to lines of sight passing near the cloud surface. The value $b_\mathrm{frac} = 0$ recovers the slab models.
    }
    \label{fig:BeckerSiIIvsCII}
\end{figure*}

\begin{figure*}
    \centering
    \includegraphics[trim=2cm 1.8cm 4.5cm 3cm,clip, width=\linewidth]{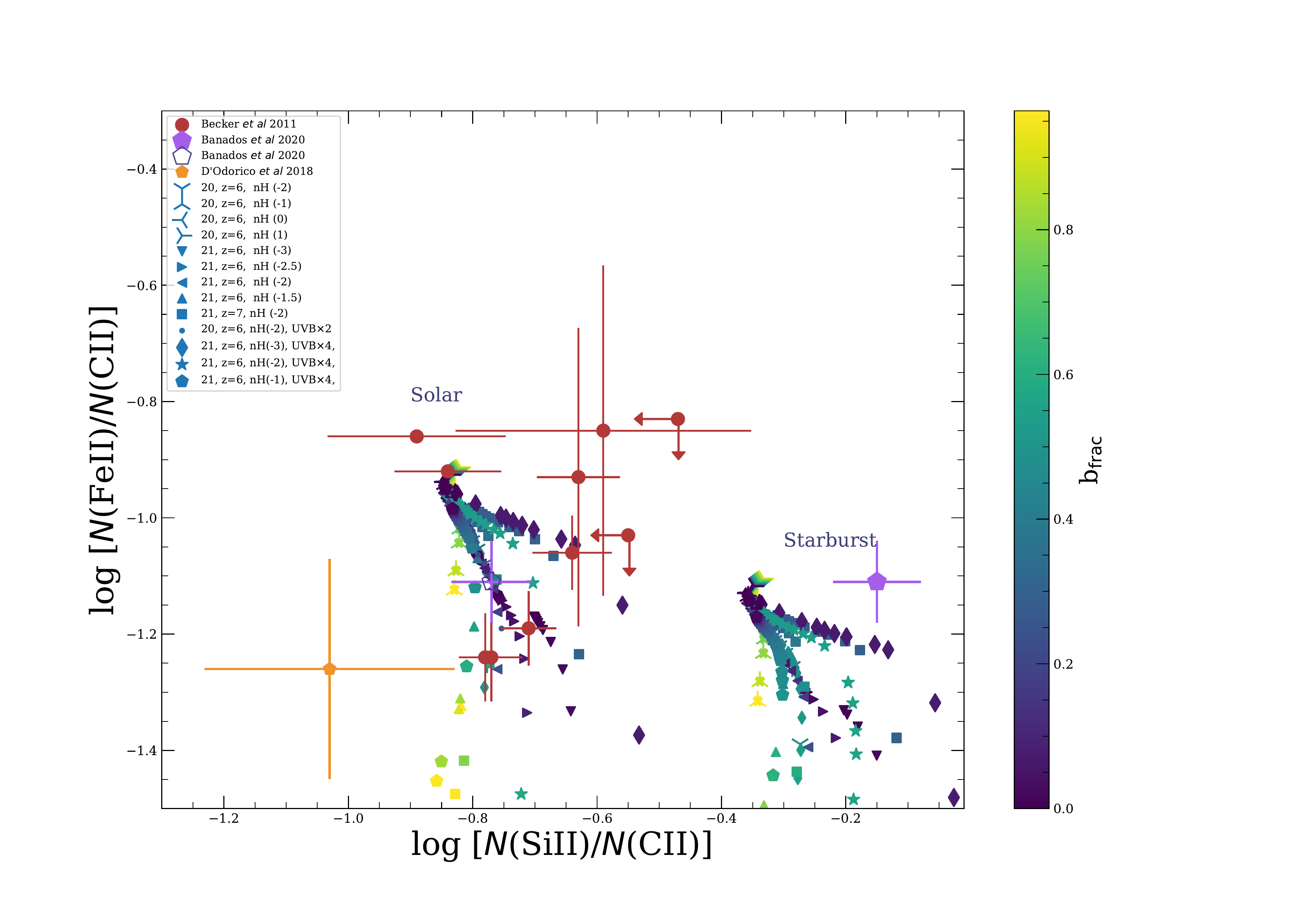}
\caption{Column density ratios between Fe{\sc ii}, Si{\sc ii}, C{\sc ii}  for data from \citet{Becker2011-12} (red circles), \citet{Banados2019ApJ...885...59B} (full and empty pink pentagons; the Si{\sc ii} values are upper limits, see text) and \citet{D_Odorico_2018} (orange pentagon). Markers in colour correspond to different {\sc cloudy} models, indicated in the legend as in Fig.~\ref{fig:BeckerSiIIvsCII}.
}
    \label{fig:BeckerSiIItoCIIvsFeIItoCII}
\end{figure*}

\begin{table*}
    \centering
\caption{Top table: Column densities and their ratios (both expressed as log), for representative ionisation models using {\sc cloudy} to model the data from \citet{Becker2011-12, Becker:2011}. (The metal column densities are normalised to $0.001Z_{\odot, Si}$ for both solar and starburst abundance models.)  A subsample of the data in Figs.~\ref{fig:BeckerSiIIvsCII} and \ref{fig:BeckerSiIItoCIIvsFeIItoCII} is shown. Bottom table: The physical properties and parameters of the models: hydrogen density  ($n_\mathrm{H}$) in cm$^{-3}$, pressure ($P$) in dyne cm$^{-2}$, Jeans length ($\lambda_\mathrm{J}$) in kpc, radius of the cloud  ($R_c)$ in kpc , mass of the cloud  ($M_c$) in $M_\odot$, impact parameter  ($b_\perp$),   $b_\text{frac}=b_\perp/R_c$,  radiation field redshift,  abundance and input H{\sc i} density for {\sc cloudy} (log $N$(H{\sc i})). The abundance column indicates starburst abundances (Sb) or solar abundances ($\odot$).
}
\begin{tabular}{rccccccccccccc}
\hline
Model &   H{\sc i} &   C{\sc ii} &    O{\sc i} &  Si{\sc ii} &  Fe{\sc ii} &   C{\sc iv}  &  Si{\sc iv} &  C{\sc ii}/O{\sc i} &  Si{\sc ii}/O{\sc i} &  C{\sc iv} /O{\sc i} &  Si{\sc iv}/O{\sc i} &  Fe{\sc ii}/C{\sc ii} &  Si{\sc ii}/C{\sc ii} \\ 
   
   \hline

1&  20.30 & 13.34 & 14.11 & 13.04 & 12.12 & 12.39 & 12.69 & -0.76 & -1.06 & -1.71 & -1.42 &  -1.22 &  -0.29 \\
2&  20.30 & 13.27 & 14.12 & 12.92 & 12.10 &  9.67 & 10.45 & -0.85 & -1.19 & -4.44 & -3.66 &  -1.16 &  -0.34 \\
3&  17.82 & 11.56 & 11.71 & 11.22 & 10.24 &  6.46 &  7.42 & -0.15 & -0.49 & -5.24 & -4.29 &  -1.31 &  -0.34 \\
4&  20.27 & 13.82 & 13.96 & 13.03 & 12.78 & 12.86 & 12.67 & -0.13 & -0.92 & -1.09 & -1.28 &  -1.04 &  -0.79 \\
5&  17.52 & 12.30 & 11.05 & 11.38 & 10.54 &  9.99 & 10.28 &  1.24 &  0.33 & -1.05 & -0.76 &  -1.76 &  -0.91 \\
6&  20.30 & 13.76 & 13.99 & 12.92 & 12.78 & 10.14 & 10.43 & -0.23 & -1.06 & -3.84 & -3.55 &  -0.97 &  -0.83 \\
7&  20.07 & 13.49 & 13.76 & 12.64 & 12.55 &  6.76 &  7.16 & -0.27 & -1.12 & -7.00 & -6.60 &  -0.94 &  -0.84 \\
8&  21.29 & 14.20 & 15.11 & 13.85 & 13.07 & 11.89 & 12.18 & -0.90 & -1.26 & -3.21 & -2.92 &  -1.13 &  -0.35 \\
9&  20.15 & 13.11 & 13.98 & 12.77 & 11.96 &  8.91 &  9.61 & -0.86 & -1.20 & -5.06 & -4.36 &  -1.14 &  -0.34 \\

\hline
\end{tabular}

\begin{tabular}{rcccccccccc}
\hline
Model  &  $n_\mathrm{H}$ &  log $P$ & $\lambda_\mathrm{J}$ &  $R_c $ & log  $M_c$ &  $b_\perp$ &  $b_\text{frac}$ & redshift & abundance & log $N$(H{\sc i}) \\
\\
\hline
\hline
1&  0.001 & -14.25 &  41.77 &  149.93 &    11.88 &   0.00 & 0.00 &   6 &    Sb  &   20 \\
2&  0.01  & -13.37 &  11.42 &    2.03 &     7.61 &   0.03 & 0.01 &   6 &    Sb  &   20 \\
3&  0.1   & -12.50 &   3.23 &    0.04 &     3.58 &   0.04 & 0.98 &   6 &    Sb  &   20 \\
4&  0.001 & -14.24 &  41.84 &  146.07 &    11.85 &   2.13 & 0.01 &   6 &  $\odot$ &   20 \\
5&  0.01  & -13.37 &  11.43 &    1.98 &     7.58 &   1.94 & 0.97 &   6 &  $\odot$ &   20 \\
6&  0.01  & -13.37 &  11.43 &    1.98 &     7.58 &   0.03 & 0.01 &   6 &  $\odot$ &   20 \\
7&  0.1   & -12.49 &   3.23 &    0.04 &     3.46 &   0.00 & 0.00 &   6 &  $\odot$ &   20 \\
8&  0.001 & -14.29 &  39.88 &  128.38 &    12.13 &   4.30 & 0.03 &   7 &    Sb  &   21 \\
9&  0.01  & -13.41 &  11.01 &    0.97 &     6.52 &   0.00 & 0.00 &   7 &    Sb  &   21 \\

\hline
\end{tabular}
    \label{ModelsBecker:table}
\end{table*}

\begin{figure*}
    \centering
    \includegraphics[ width=\linewidth]{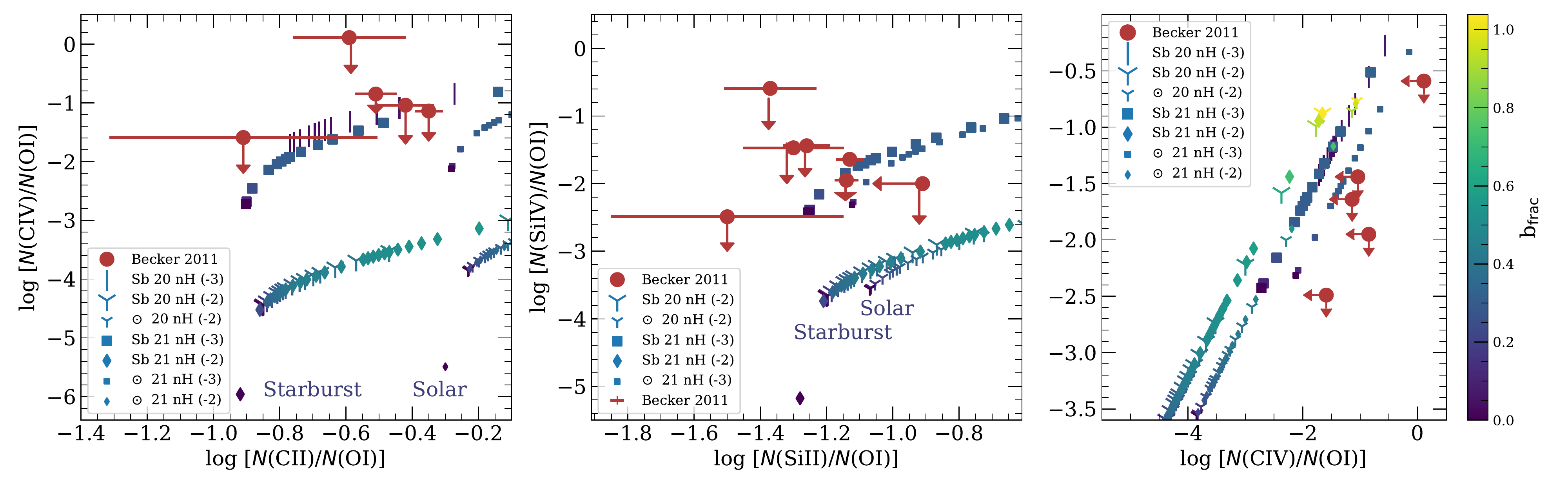}
    \caption{
    Data taken from \citet{Becker:2011}, all upper limits. Coloured markers indicate several combinations of parameters of the {\sc cloudy} models.  See properties in the legend box. There is a near degeneracy between solar and starburst abundances for some ratios.  The colour bar indicates the impact parameter of the line-of-sight measurement through the cloud, with dark blue corresponding to small impact parameters and light yellow to lines of sight passing near the cloud surface. The value $b_\mathrm{frac} = 0$ recovers the slab models.
(left panel) Column densities of C{\sc iv} to OI vs C{\sc ii} to O{\sc i}. The values in the column densities for C{\sc iv} systems are upper limits (red arrows).
    (middle panel)
    Column densities for of Si{\sc iv} to O{\sc i} vs Si{\sc ii} to O{\sc i}.
 Arrows indicate upper limits for Si{\sc iv}; one data point indicates a lower limit in O{\sc i}.  
(right panel) Column densities systems of Si{\sc iv} vs C{\sc iv} (both to O{\sc i}).
    }
    \label{fig:total_SiIICIISiIVCIV}
\end{figure*}

\section{Comparison with observations}
\label{observations}   

We use measurements of the low ionisation species of carbon, oxygen, silicon and iron in intervening absorption systems at $z\gtrsim5$ from \citet{Becker:2011, Becker2011-12, DOdorico2013MNRAS.435.1198D, RyanWeber2009MNRAS.395.1476R, 2016ApJ...830..158M, Poudel2018, Poudel2019, Banados2019ApJ...885...59B} to constrain the physical properties of the absorbers.
We provide a compilation of the data used in Appendix~\ref{obs:tables}: Tables~\ref{Becker:Table3}, \ref{Poudel:table01}, \ref{Poudel:tables02} and \ref{Dodorico}.

We first compare with the data of \citet{Becker2011-12}. The metal line measurements were taken using the Keck/HIRES (FWHM = 6.7 km s$^{-1}$), Keck/NIRSPEC (FWHM $\sim$ 15 km s$^{-1}$), Magellan/MIKE (FWHM = 13.6 km s$^{-1}$) and VLT/X-Shooter ($\sim 30$ km s$^{-1}$) spectrometers. We also compare with two very low metallicity ($Z\sim0.001Z_\odot$) proximate DLA systems at $z\gtrsim6$. One measured in the spectrum of the QSO J2310 using the X-Shooter spectrometer with a resolution of 8800 has redshift $z\simeq5.939$, blueshifted relative to the systemic QSO redshift by $\sim2750$~km~s$^{-1}$ with log N(H{\sc i}) $= 21.05\pm0.1$ \citep{D_Odorico_2018}. The second system was detected in the spectrum of QSO P183$+$05 that was selected in the Pan-STARRS1 survey \citep{PANSTARRS2016arXiv161205560C} at $z=6.4$ \citep{Banados2019ApJ...885...59B}. The absorber is blueshifted relative to the QSO systemic redshift by only 1400~km~s$^{-1}$, or at a distance of 1.8~Mpc in front of the QSO if the velocity is interpreted as Hubble expansion. The H{\sc i} column density measured in the QSO spectrum exceeds $10^{20.5}\,{\rm cm}^{-2}$, but it is unclear how much of this arises in the absorption system and how much from the ambient IGM, especially as the IGM may not be fully ionised at this redshift. The data are also consistent with the system being a sub-DLA in a mostly neutral IGM.

Motivated by previous findings that the column density ratios among C{\sc ii}, Si{\sc ii}, O{\sc i} and Fe{\sc ii}  are consistent with (metal poor) sub-DLAs and DLAs \citep{Becker:2011, Becker2011-12}, all our models use neutral hydrogen column densities of $N$(H{\sc i}) $\sim10^{20} - 10^{21}\,{\rm cm}^{-2}$, to represent the systems. We also explored the effect of using lower column densities, log $N$(H{\sc i})$\lesssim19$, but we find that the higher column densities improve the agreement with the observed ratios. Results for both solar and starburst chemical abundances for models matching the data are provided in Table~\ref{ModelsBecker:table}.

Fig.~\ref{fig:BeckerSiIIvsCII} shows the column density ratios of the ions C{\sc ii} to O{\sc i} and Si{\sc ii} to O{\sc i}, for absorption systems at $4.7 < z < 6.3$. The colour bar indicates the impact parameter for the assumed line of sight, expressed as a fraction of the cloud radius (R$_c$):\ $b_\text{frac}=b_\perp/R_c$ (see Fig. \ref{fig:SlabCircle}). Purple corresponds to a line of sight close to the centre of the cloud and yellow is near the surface of the cloud. The value $b_\mathrm{frac} = 0$ recovers the result for a line of sight perpendicular to the corresponding slab model (except the H{\sc i} column density is doubled to account for the full cloud diameter). The models with solar abundance hit a floor in log [$N$(C{\sc ii})/$N$(O{\sc i})] at about $\sim -0.3$, corresponding to a line of sight passing through the centre of the DLA, where all the carbon is in the form of C{\sc ii} and the oxygen in the form O{\sc i}. This is because radiation able to ionise neutral oxygen or singly ionised carbon is no longer able to penetrate to the centre of the cloud. Almost all the data of \citet{Becker:2011} have smaller values. Whilst a broad range of larger ratios are possible for lines of sight at increasing impact parameters, these models move away from the measured values. Models are able to recover the measured ratios only by increasing the carbon to oxygen ratio, confirming the conclusion of \citet{Becker:2011} that the composition of the clouds must be $\alpha$-enhanced. Fig.~\ref{fig:BeckerSiIIvsCII}, however, shows it is not necessary to move all the way to the chemical composition of a starburst:\ the clouds display intermediate compositions.

The proximate DLA reported by \citet{D_Odorico_2018} shows several metal absorption lines, however they remark the lines are not resolved by X-Shooter and so may be saturated. The reported column densities for C{\sc ii}, Si{\sc ii} and O{\sc i} none the less place the absorber near the trend for solar abundances, corresponding to values deep within a DLA core (orange point in Fig.~\ref{fig:BeckerSiIIvsCII}).

The ratio of C{\sc ii} to O{\sc i} for the proximate DLA from \citet{Banados2019ApJ...885...59B} is consistent with both solar and starburst abundances, but only for impact parameters passing through the outer half of a cloud if starburst abundances are assumed and a standard UVB intensity adopted. Such a possibility is consistent with the alternative sub-DLA interpretation of the measured absorption. The reported Si{\sc ii} to O{\sc i} ratio places it squarely in the starburst abundance regime, distinct from the solar, as shown by the pink filled point in Fig.~\ref{fig:BeckerSiIIvsCII}. Boosting the UVB by a factor of 4, as if there were local sources, would permit the line of sight to pass closer to the cloud core for $\log N$(H{\sc i})$=20\,{\rm cm}^{-2}$. There are, however, discrepancies between the three distinct Si{\sc ii} transitions detected, likely a result of contamination of some of the features. The authors suggest adopting the lower column density measured of $10^{13.5}\,{\rm cm}^{-2}$, and even accepting it conservatively as an upper limit because of possible C{\sc iv} contamination from another absorption system along the line of sight. We show this upper limit in Fig.~\ref{fig:BeckerSiIIvsCII} with the empty pink pentagon-type point. This decreases log[N(Si{\sc ii})/N(O{\sc i})] to $<-0.9$, in agreement with solar abundances, yet again preferentially for a line of sight passing outside a cloud core, where the H{\sc i} and O{\sc i} fractions have fallen, although the large error bars are consistent with a line of sight passing through a cloud core as well. The errors are marginally ($\sim2\sigma$) consistent with starburst abundances.

Similar trends are found on comparing model predictions for the Fe{\sc ii} to C{\sc ii} and Si{\sc ii} to C{\sc ii} ratios with the data, as shown in Fig.~\ref{fig:BeckerSiIItoCIIvsFeIItoCII}. The absorption systems from \citet{Becker2011-12} again lie between the predictions for absorption within DLA cores with abundances between solar and starburst, although some, with the lower values of N(Fe{\sc ii})/ N(C{\sc ii}), favour lines of sight with impact parameters offset from the cores. Allowing for an enhanced radiation field also introduces an ambiguity in interpretation, producing matches to the data with higher values of N(Fe{\sc ii})/ N(C{\sc ii}) if along lines of sight displaced from the cloud cores for systems with solar abundances. This is opposite the trend for the Si{\sc ii} to O{\sc i} and C{\sc ii} to O{\sc i} ratios, suggesting intermediate abundances for the absorbers is the preferred interpretation.

Given the large Fe{\sc ii} error bar for the proximate DLA of \citet{D_Odorico_2018} (orange point), and the possibility the C{\sc ii} line is saturated, the system is again consistent with a line of sight through the core of a DLA with solar abundances. The Fe{\sc ii} to C{\sc ii} and Si{\sc ii} to C{\sc ii} ratios for the proximate DLA of \citet{Banados2019ApJ...885...59B}, accepting the larger Si{\sc ii} column density, is again consistent with a line of sight displaced from the core of an absorber with starburst abundances. Accepting the lower Si{\sc ii} as an upper limit, however, moves the point to log[N(Si{\sc ii})/N(C{\sc ii})] $< -0.77$, in agreement with a sub-DLA interpretation with solar abundances, as above for the Si{\sc ii} to O{\sc i} and C{\sc ii} to O{\sc i} ratios.

Constraints from the upper limits placed on high-ionisation ions by \citet{Becker:2011} are shown in Fig.~\ref{fig:total_SiIICIISiIVCIV}. These provide consistency checks on the models inferred from the low-ionisation line ratios. The left and middle panels show the column density ratios of C{\sc iv} and Si{\sc iv}, each to O{\sc i}, vs the column density ratios of the low-ionisation lines C{\sc ii} and Si{\sc ii}, each to O{\sc i}. The high ionisation lines place an additional constraint on the density of the absorption systems, requiring $\log n_{\rm H}>-3$. Combining the C{\sc iv} and Si{\sc iv} upper limits is consistent with this, as shown in the third panel.


\begin{figure*}
    \centering
    \includegraphics[width=\linewidth]{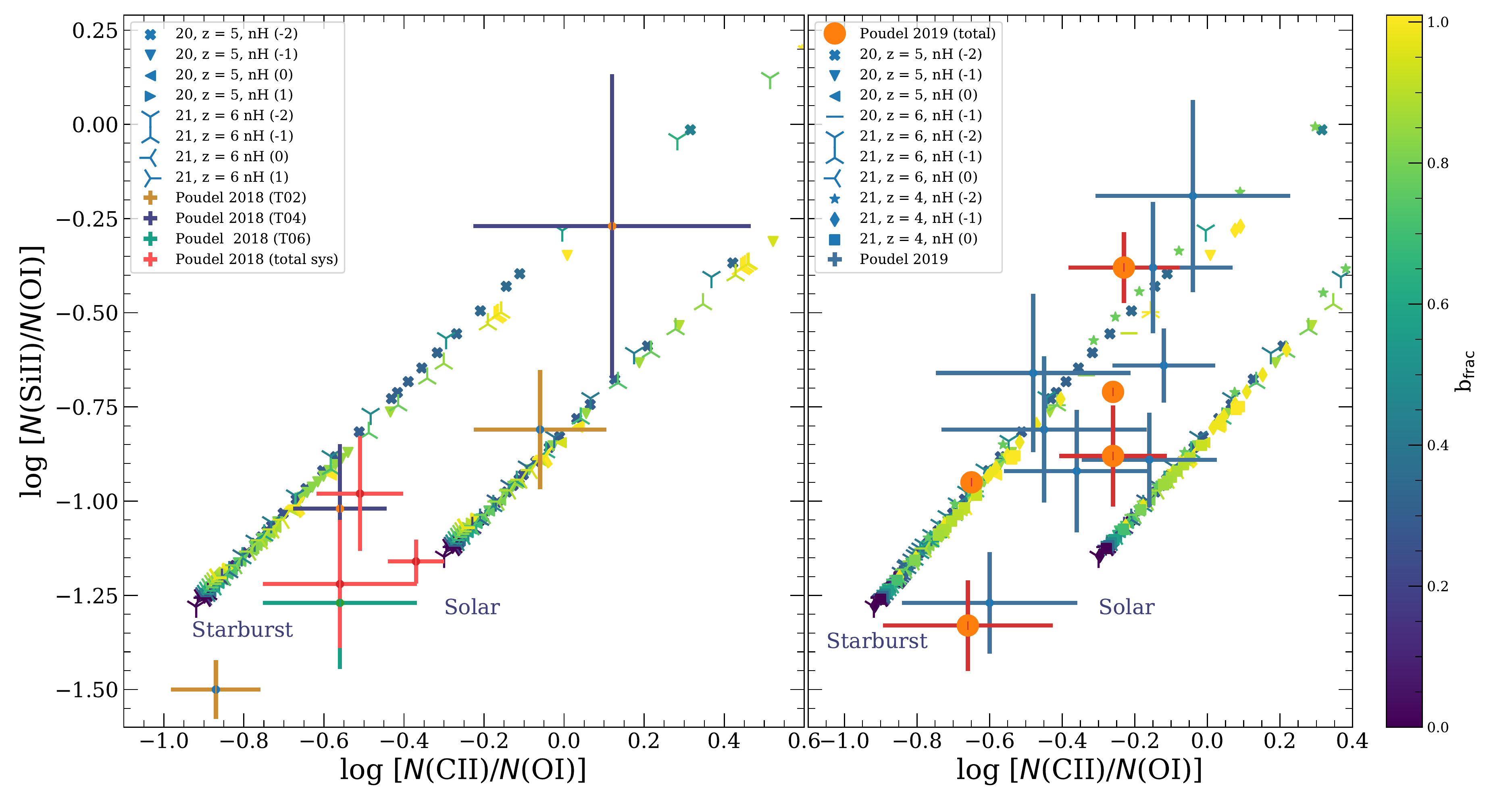}
    \includegraphics[trim=1.43cm 1.0cm 1.68cm 1cm,clip,width=\linewidth]{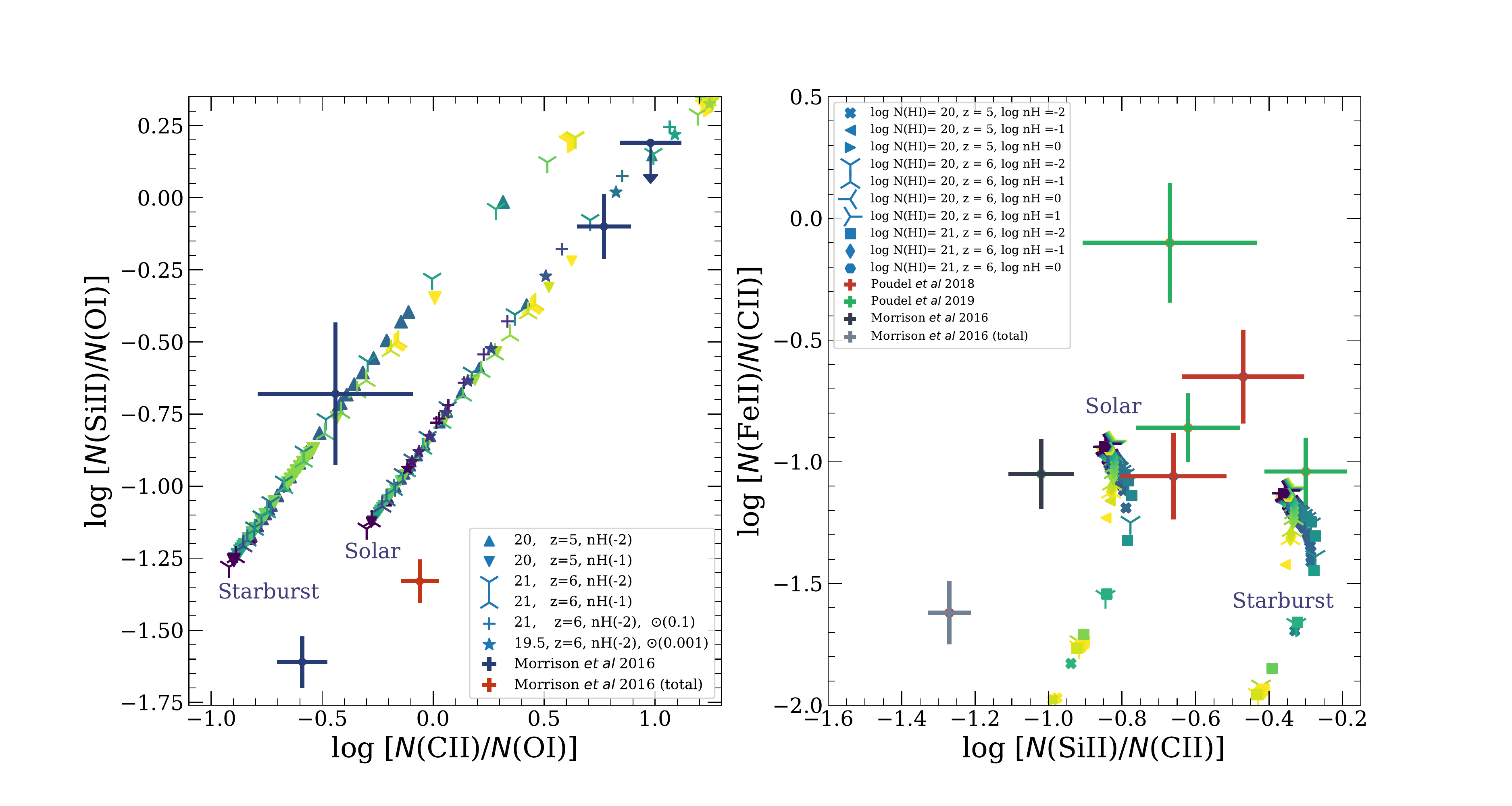}
    \caption{
(Top left panel): Column density ratios of C{\sc ii} to O{\sc i} vs Si{\sc ii} to O{\sc i}. Data from \citet{Poudel2018}. Red crosses indicate absorption complexes; blue crosses are for individual sub-components. Cloud models using {\sc cloudy} are also shown, with the legends listing $\log N$(H{\sc i}) of the corresponding slab model, the redshift of the UVB model, and the surface hydrogen density. (Top right panel): Column density ratios for ion C{\sc ii} to O{\sc i} vs Si{\sc ii} vs O{\sc i}. The blue points/crosses correspond to distinct sub-components observed at $z>4.5$. The orange points/red crosses are the total systems. Data taken from \citet{Poudel2019}. Results for {\sc cloudy} models are also shown.
(Bottom left panel): Column density ratios for Si{\sc ii} to O{\sc i} vs C{\sc ii} to O{\sc i}. Data from \citet{2016ApJ...830..158M}. Red marker is a total system. The system on the top right is for an upper limit on Si{\sc ii}.
(Bottom right panel): Column density ratios for Fe{\sc ii} to C{\sc ii} vs Si{\sc ii} to C{\sc ii}. These are the total systems for data from both \citet{Poudel2018} and \citet{Poudel2019}. We have included one individual sub-component and a total system from \citet{2016ApJ...830..158M}. For all panels, the colours for the {\sc cloudy} models indicate the impact parameter of the line of sight, as indicated by the colour bars. The value $b_\mathrm{frac} = 0$ recovers the slab models.
    }
    \label{fig:totalPoudel}
\end{figure*}

\begin{table*}
    \centering
\caption{Top table: Column densities and their ratios (both expressed as log), for representative ionisation models using {\sc cloudy} to model the data in Fig.~\ref{fig:totalPoudel} from \citet{Poudel2018} and \citet{Poudel2019}. Bottom panel: Physical properties and parameters for the models: hydrogen density ($n_\mathrm{H}$) in cm$^{-3}$, pressure ($P$) in dyne cm$^{-2}$, Jeans length ($\lambda_\mathrm{J}$) in kpc, radius of the cloud  ($R_c)$ in kpc, mass of the cloud  ($M_c$) in $M_\odot$, impact parameter  ($b_\perp$) in kpc, $b_\text{frac}$,  model radiation field redshift,  abundance and input H{\sc i} density for {\sc cloudy} (log $N$(H{\sc i})). The abundance column indicates starburst abundances (Sb) or solar abundances ($\odot$). (The metal column densities are normalised to $0.001Z_{\odot, Si}$ for both solar and starburst abundance models.)
}
\begin{tabular}{rccccccccc}
\hline
 Model &  H{\sc i} &   C{\sc ii} &    O{\sc i} &  Si{\sc ii} &    Fe{\sc ii}     & C{\sc ii}/O{\sc i} &  Si{\sc ii}/O{\sc i} &  Fe{\sc ii}/C{\sc ii} &  Si{\sc ii}/C{\sc ii} \\
\hline
 1&  19.03 & 12.77 & 12.88 & 12.48 & 11.36 & -0.11 & -0.39 &  -1.41 &  -0.28 \\    
 2&  20.16 & 13.19 & 13.98 & 12.85 & 11.99 & -0.79 & -1.13 &  -1.19 &  -0.34 \\    
 3&  19.95 & 12.93 & 13.78 & 12.58 & 11.78 & -0.84 & -1.20 &  -1.14 &  -0.35 \\    
 4&  20.26 & 13.18 & 14.08 & 12.82 & 12.04 & -0.89 & -1.25 &  -1.13 &  -0.35 \\    
 5&  20.30 & 13.22 & 14.12 & 12.86 & 12.08 & -0.89 & -1.25 &  -1.13 &  -0.35 \\    
  6&  20.30 & 13.78 & 13.99 & 12.94 & 12.79 & -0.21 & -1.05 &  -0.99 &  -0.84 \\    
7& 19.71 & 13.45 & 13.42 & 12.64 & 12.38 &  0.03 & -0.78 &  -1.07 &  -0.81 \\
8& 19.90 & 13.54 & 13.61 & 12.71 & 12.49 & -0.07 & -0.90 &  -1.04 &  -0.82 \\
 9&  20.04 & 13.49 & 13.73 & 12.64 & 12.53 & -0.24 & -1.09 &  -0.95 &  -0.84 \\    
 10&  20.26 & 13.67 & 13.95 & 12.83 & 12.73 & -0.27 & -1.12 &  -0.94 &  -0.84 \\    
 11&  19.41 & 12.83 & 13.27 & 12.54 & 11.57 & -0.44 & -0.72 &  -1.25 &  -0.28 \\    
12&  20.31 & 13.28 & 14.13 & 12.94 & 12.12 & -0.84 & -1.19 &  -1.16 &  -0.34 \\    
13&  19.19 & 12.34 & 13.04 & 12.01 & 11.19 & -0.69 & -1.03 &  -1.15 &  -0.33 \\    
14&  19.60 & 12.58 & 13.43 & 12.23 & 11.44 & -0.84 & -1.19 &  -1.14 &  -0.35 \\ 

\hline
\end{tabular}
\begin{tabular}{rcccccccccc}
\hline
Model &  $n_\mathrm{H}$ &  log $P$ & $\lambda_\mathrm{J}$ &  $R_c $ &  log $M_c$ &  $b_\perp$ &  $b_\text{frac}$ & redshift & abundance & log $N$(H{\sc i}) \\
\hline

 1&  0.01 & -13.35 &  11.71 &   2.83 &     7.96 &   0.97 & 0.34 &    5 &    Sb  &       20 \\
 2&  0.01 & -13.35 &  11.71 &   2.83 &     7.96 &   0.53 & 0.18 &    5 &    Sb  &       20 \\
 3&  0.1  & -12.47 &   3.28 &   0.12 &     4.93 &   0.08 & 0.67 &    5 &    Sb  &       20 \\
 4&  0.1  & -12.47 &   3.28 &   0.12 &     4.93 &   0.03 & 0.29 &    5 &    Sb  &       20 \\
 5&  0.1  & -12.47 &   3.28 &   0.12 &     4.93 &   0.00 & 0.00 &    5 &    Sb  &       20 \\
 6&  0.01 & -13.34 &  11.72 &   2.77 &     7.93 &   0.00 & 0.00 &    5 &  $\odot$  &    20 \\
7    & 0.01  & -13.34 & 11.72 &   2.77 &   7.93 &  0.78 & 0.28&  5 &  $\odot$ &  20 \\
8& 0.01  & -13.34 & 11.72 &   2.77 &   7.93 &  0.72 & 0.26&  5 &  $\odot$ &  20 \\   
 9&  0.1  & -12.46 &   3.28 &   0.12 &     4.91 &   0.07 & 0.62 &    5 &  $\odot$  &    20 \\
 10&  0.1  & -12.46 &   3.28 &   0.12 &     4.91 &   0.03 & 0.29 &    5 &  $\odot$  &    20 \\
 11&  0.01 & -13.38 &  11.43 &   2.24 &     7.77 &   1.14 & 0.51 &    6 &    Sb  &       21 \\
12&  0.01 & -13.38 &  11.43 &   2.24 &     7.77 &   0.69 & 0.31 &    6 &    Sb  &       21 \\
13&  0.1  & -12.50 &   3.23 &   0.04 &     3.58 &   0.02 & 0.63 &    6 &    Sb  &       21 \\
14&  0.1  & -12.50 &   3.23 &   0.04 &     3.58 &   0.01 & 0.32 &    6 &    Sb  &       21 \\

    \hline
\end{tabular}
    \label{ModelsPoudel:table}
\end{table*}

\citet{Poudel2018} targetted three DLA systems at $4.8<z<5.4$, somewhat later in cosmic time than the systems above but still probing into the cosmic dawn era.

A summary of the observations is provided in Table~\ref{Poudel:table01}, which shows the results from Voigt-profile fitting for three absorption systems at $z_{\rm abs}=5.335$ (SDSS QSO Q0231$\pm$0728), $z_{\rm abs} = 4.809$ and 4.829 (Q0824+1302). All these systems show low-ionisation metal absorption features of C{\sc ii} and Si{\sc ii} and O{\sc i}. We also include an earlier measurement of a sub-DLA at $z=4.98$ in the spectrum of SDSS QSO Q1202$+$3235 using the Keck HIRES and ESI spectrometers \citep{2016ApJ...830..158M}.

These samples were enhanced by another set of systems at $z>4.5$, some now measured at very high spectral resolution \citep{Poudel2019}. The two quasars J1557$+$1018 and J1253$+$1046 were observed with MIKE at the Las Campanas observatory,  QSO J1233$+$0622 was observed with VLT X-Shooter and QSO J0306$+$1853 was observed with MagE (Magellan Echellette). The spectral resolution for MIKE is $\sim$ 22,000-28,000, for X-Shooter is R = 5300 and for MagE is $\sim4100$; \citep[see][for details.]{Poudel2019}.
 
The sample contains seven absorbers with neutral hydrogen column density of $\log N$(HI)=19.65 to 20.75, at redshifts from z = 4.59 to 5.05 along the sight lines to four quasars. Table~\ref{Poudel:tables02} presents the results of Voigt-profile fitting for O{\sc i}, C{\sc ii}, Si{\sc ii} and Fe{\sc ii}.
This is a subsample of all the reported measurements in Table 2 in \citet{Poudel2019}, showing C{\sc ii} and Si{\sc ii} and at least one of O{\sc i} or Fe{\sc ii}.

In Fig.~\ref{fig:totalPoudel} , we compare estimates for the column density ratios C{\sc ii} to O{\sc i} and Si{\sc ii} to O{\sc i} from different models, following the procedure described for Figs.~\ref{fig:BeckerSiIIvsCII} and \ref{fig:BeckerSiIItoCIIvsFeIItoCII}, with the data from \citet{2016ApJ...830..158M} and \citet{Poudel2018, Poudel2019}. We again consider both solar and starburst abundances. The top-left panel shows results for three absorption complexes observed at moderate spectral resolution that Voigt-profile fitting resolves into a few sub-components. The ratios drawn from the total column densities for the complexes are shown as red points. We next discuss these complexes and their components.

The absorber at $z_\text{abs}=5.335$ is fit by two sub-components \citep[yellow points; data taken from Table 2 in][]{Poudel2018}. One individual sub-component from this dataset is matched by models with solar abundances (such as Model 8 in Table~\ref{ModelsPoudel:table}). The other sub-component lies closer ($\sim 3\sigma$) to the models using starburst abundances, however it still has an offset of $\sim$ 0.2 dex from the starburst models and no model provides a good fit:\ the Si{\sc ii} to O{\sc i} ratio is too low. Using the
total column densities for this system, however, the ratios agree with a solar abundance model with $\log n_\mathrm{H} = 1\,{\rm cm}^{3}$ (red point with the smallest error bars). We are not able to push the hydrogen density to higher values, as {\sc cloudy} stops the calculation because the electron temperature reaches its lowest possible value for an isobaric cloud. 
The absorption complex illustrates that insufficient resolution may obscure distinct physical differences in sub-components.

The absorption system at $z_\text{abs}=4.809$ \citep[purple points; data from Table 4 in][]{Poudel2018} is matched by the models as well as its parent total system, which differs little from the sub-component with the smaller error bars. This latter is best fit by a model with starburst abundances. The other component is matched by both starburst and solar abundances, but for a line of sight well offset from the cloud centre, although the error bars are large.

Finally, both solar and starburst abundance models are consistent with the system at $z_\text{abs}=4.829$ \citep[green point; taken from Table 6 in][]{Poudel2018} because of its large error bars. The Si{\sc ii} to O{\sc i} ratio prefers lines of sight passing through or near a cloud centre.


The top-right panel in Fig.~\ref{fig:totalPoudel} shows the column density ratios for C{\sc ii} to O{\sc i} vs Si{\sc ii} to O{\sc i} from \citet{Poudel2019}
(a summary of the data is in \ref{Poudel:tables02}).
As for the data of \citet{Poudel2018}, metal absorption complexes are Voigt-profile fit with sub-components. We note that the absorption systems at $z_\text{abs}<4.8$ were observed at much higher spectral resolution than the higher redshift absorbers or any of the systems in \citet{Poudel2018}. We model the data using the same metallicity models described above for \citet{Poudel2018} (Table~\ref{ModelsPoudel:table}).

The models with starburst abundances generally better match the data than solar abundances models for the resolved metal sub-components. Lines of sight passing through the centres of the clouds, however, are often in poor agreement with the data, although lines of sight passing through the inner third of the lower H{\sc i} column density cloud models are acceptable. Alternatively, they may arise from lines of sight passing through the outer half of a high density model (such as Model 9 in Table~\ref{ModelsPoudel:table}). The ratios based on the total metal column densities are more ambiguous regarding the abundances. Only the system at $z_\text{abs}=4.987$ gives a clear preference for starburst abundances. Two other systems (shown as points without error bars) are indeterminate, as the total column densities are provided as formal lower limits.

Results for the sub-DLA system at $z_\text{abs}=4.98$ from \citet{2016ApJ...830..158M} is shown in the bottom left panel in Fig.~\ref{fig:totalPoudel}. Using the total metal column densities for the system strongly favours solar abundances, but requires silicon depletion at a level $0.15-0.4$~dex. The sub-components from Voigt-profile fitting the feature, however, are more ambiguous regarding the abundances. Only one sub-component clearly favours solar abundances, and also requires the line of sight to pass outside a cloud core. A second lies between solar and starburst abundances and is consistent with both, but the Si{\sc ii} to O{\sc i} ratio requires substantial depletion by $0.3-0.4$~dex. The depletion analysis by \citet{2016ApJ...830..158M} confirms this as a high depletion system. A third sub-component favours starburst abundances, but its large error bars permit consistency with solar as well. The complex also has detected C{\sc iv} and Si{\sc iv} absorption, but given the small errors on the redshifts of the sub-components ($\Delta z_\text{abs}<10^{-4}$), it is unclear these sub-components should be associated with the low-ionisation sub-components, or to which sub-component, so we refrain from comparisons with the models.

The bottom right panel in Fig.~\ref{fig:totalPoudel} shows Fe{\sc ii} systems. We show ratios for the total column densities of Fe{\sc ii} to C{\sc ii} vs Si{\sc ii} to C{\sc ii} from absorption complexes \citep[compiled from both][]{Poudel2018, Poudel2019}. The data for individual sub-components for Fe{\sc ii} is incomplete, with Fe{\sc ii} undetected in any sub-component in the \citet{Poudel2018} data for which O{\sc i}, C{\sc ii} or Si{\sc ii} was detected. Only one sub-component, at $z_\text{abs}=5.05024$, in the data from \citet{Poudel2019} had a detection in all three of C{\sc ii}, Si{\sc ii} and Fe{\sc ii}. Given the large errors, the data are generally consistent with either solar or starburst abundances, for lines of sight passing preferentially through the cloud centres, but the data do have generally high values of Fe{\sc ii} to C{\sc ii}, especially the system at $z_\text{abs}=4.589$. The sub-DLA at $z_\text{abs}=4.98$ from \citet{2016ApJ...830..158M}, however, now clearly favours solar abundances, with a silicon depletion factor relative to carbon of $\sim 0.3-0.5$~dex.


\begin{table*}
    \centering
\caption{Summary of the physical properties and parameters for models of the data from \citet{DOdorico2013MNRAS.435.1198D}. Listed are the hydrogen density ($n_\mathrm{H}$) in cm$^{-3}$, pressure ($P$) in dyne cm$^{-2}$,  Jeans length ($\lambda_\mathrm{J}$) in kpc, cloud radius ($R_c$) in kpc, cloud mass ($M_c$) in $M_\odot$ and parameter values for the ionisation models:\ redshift of UV background model, chemical abundance and metallicity, and H{\sc i} column densities, for models shown in Fig.~\ref{fig:dodorico}.  }
    
    \begin{tabular}{cccccccc}
\hline
\hline
 log $n_\mathrm{H}$ &  log $P$   & $\lambda_\mathrm{J}$ &    $R_c$ &   log $M_c$  & $z$&  Ab, $Z/Z_\odot$  & log $N$(H{\sc i}) \\
 

\hline

-3.00 &    -14.25 &     41.78 	&    149.94 	&       11.89 &        6 		&    Sb 0.001 	&       20 \\
-2.00 &    -13.38 &    11.43 	&      2.04	 &        7.61 &        6 		&    Sb 0.001 	&       20 \\
 -3.00 &    -14.32 &    38.19 	&    136.78	 &       11.63 &        6 		&  $\odot$  0.1 	&       20 \\
-2.00 &    -13.41 &    11.03 	&      1.34 	&        6.75 &        6 		&  $\odot$ 0.1	&       20 \\ 
 -3.00 &    -14.04 &   52.85	 &    785.48     &       13.97 &  6 UVB$\times$4	&    Sb 0.001 . &       18 \\
-2.00 &    -13.27 &     12.97 	&      8.22 	&        8.94 &  6 UVB$\times$4 	&    Sb 0.001  &       18 \\
-3.00 &    -14.17 &     45.51 	&    581.34 	&       13.48 &   6 UVB$\times$4 	&  $\odot$ 0.01  	&       18 \\
-2.00 &    -13.30 &     12.37 	&      4.50 	&        8.15 &    6 UVB$\times$4	 &  $\odot$ 0.01 	&       18 \\
-3.00 &    -14.04 &     52.27 	&    703.18 	&       13.83 & 6 HM05 UVB 4 	&  $\odot$ 0.01 	&       18 \\
-2.00 &    -13.26 &     12.91 	&      7.98 	&        8.90 & 6 HM05 UVB 4 	&  $\odot$  0.01 	&       18 \\
-3.00 &    -14.11 &     48.66 	&     467.92 	&       13.26 &  7 HM05 UVB 3	 &  $\odot$  0.001 	&       18 \\
-2.00 &    -13.29 &     12.44 	&      4.37 	&        8.11 &  7 HM05 UVB 3 	&  $\odot$  0.001 	&       18 \\
\hline
\end{tabular}
    \label{tab:dodorico}
\end{table*}


\begin{figure}
    \centering  
    \includegraphics[trim=2cm 1.8cm 4.5cm 3cm,clip,width=\linewidth]{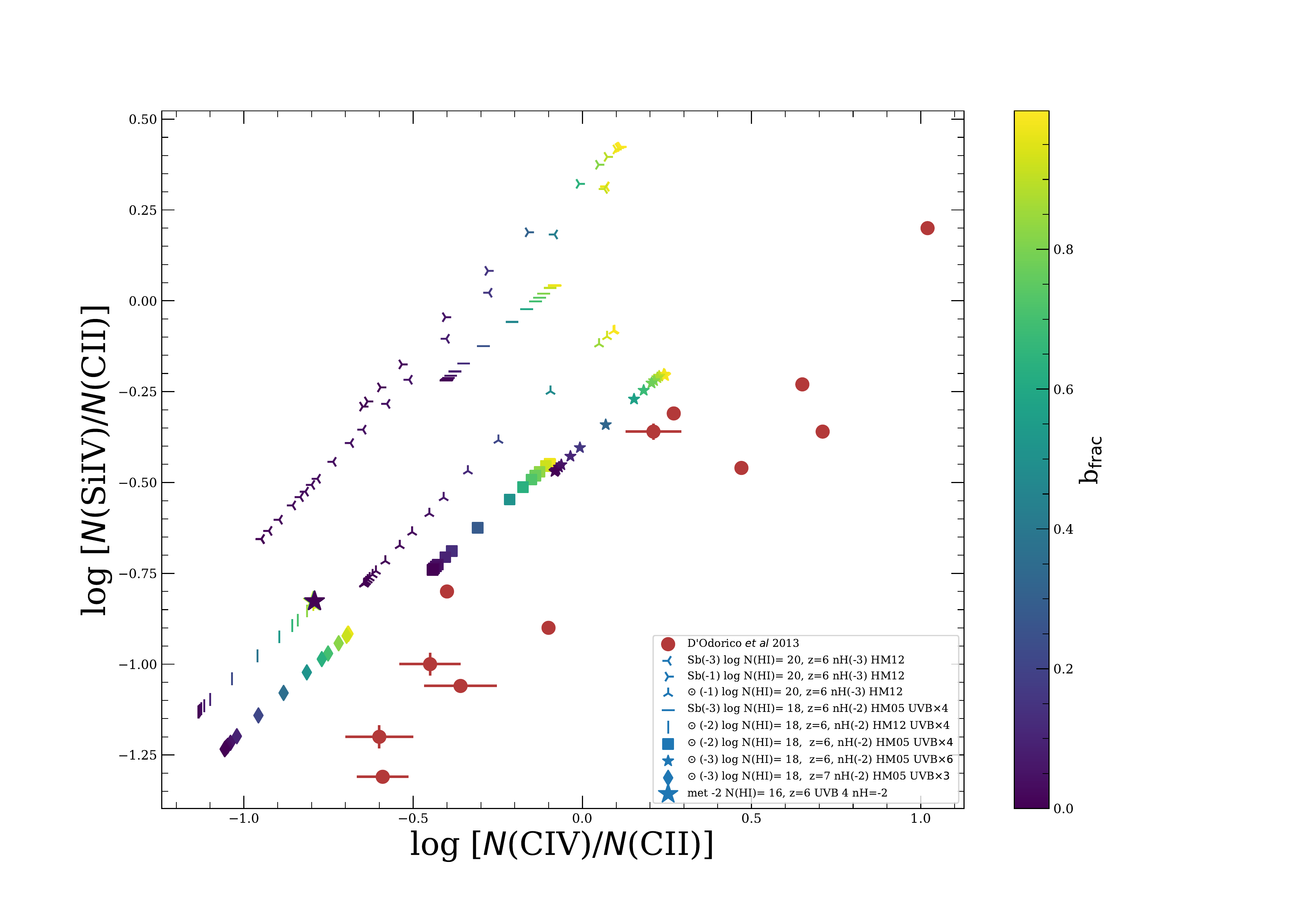}
    \caption{Column density ratios of C{\sc iv}  to C{\sc ii} vs Si{\sc iv} to C{\sc ii}. Measurements are from  \citet{DOdorico2013MNRAS.435.1198D}. Ionisation models using {\sc cloudy} are indicated by the coloured markers. The colour indicates the impact parameter of the line of sight for each model, as shown by the colour bar. The value $b_\mathrm{frac} = 0$ recovers the slab models.}
    \label{fig:dodorico}
\end{figure}

Because the metals in the outer layers of sub-DLA and DLA systems may exist in high ionisation states, we also model the high redshift data for high ionisation absorbers from \citet{DOdorico2013MNRAS.435.1198D}. Column density ratios between for C{\sc iv} , Si{\sc iv} and C{\sc ii} detected at $5<z<6$ are shown in Fig.~\ref{fig:dodorico}. These measurements were taken using the X-Shooter spectrograph with resolution between 27 - 53 km s$^{-1}$. The signal-to-noise ratios (SNRs), computed for each resolution element, range over $7<\text{SNR}<  125$. (See Table~\ref{tab:dodorico} for details.)

Isobaric cloud models using {\sc cloudy} do not provide good matches to these systems, as shown in Fig.~\ref{fig:dodorico}, but lie systematically offset from the trends. The offset may arise either because of a low silicon to carbon ratio in the observed systems, with depletion factors up to $\sim0.5$~dex, or from too low ionisation. Models that best approximate the column densities in these systems use a radiation field from the HM05 table in {\sc cloudy}, which enhances the ultraviolet background contribution from quasars compared with the model of \citet{HM2012}. (See the {\sc cloudy} documentation for details.) For isobaric models with log $N$(HI)=20 to come near to reproducing the measured values, the hydrogen density at the surface must exceed $n_\text{H} > 0.001 \text{cm}^{-3}$ with $R_c > \lambda_\mathrm{J}$ (see table \ref{tab:dodorico}), and so are Jeans unstable. We explored alternative values for the H{\sc i} column densities from log $N$(H{\sc i}) = 16 to 21 ${\rm cm}^{-2}$. The best fitting models are for lines of sight passing through the outer regions of the clouds with low central H{\sc i} column densities of $~10^{18}$ cm$^{-2}$, similar to the slab models in \citet{DOdorico2013MNRAS.435.1198D}. Some of the data points still extend to higher ratios of Si{\sc iv} to C{\sc ii} (and C{\sc iv} to C{\sc ii}) than predicted by the models.

Models with solar abundances, and metallicities between 0.001- 0.01 $Z_\odot$, are more consistent with the observations than models with starburst abundances. We also found that, even using the HM05 radiation field, further boosting in the intensity was required, as may be provided by local radiation sources. Examples with boost factors of 3, 4 and 6 are shown. Similar conclusions were reached by \citet{DOdorico2013MNRAS.435.1198D}. Few, if any, of the observed systems appear to arise from the outer regions of pressure-confined sub-DLAs or DLAs.

\begin{table*}
    \centering
        \caption{Top table: Column densities for Si{\sc ii}, Fe{\sc ii} and Al{\sc ii} obtained using the equivalent width measurements ($W_0$)[\r{A}] of unsaturated lines
from \citet{RyanWeber2009MNRAS.395.1476R}. 
Bottom table: Estimated values for the same ions, column density ratios and their physical properties from photoionisation models for isobaric clouds using {\sc cloudy}. We take as a reference the calculated column density ratios for J084035.09$+$562419.9 log[$N$(Si{\sc ii})/N(Fe{\sc ii})] = -0.34 ,  log[$N$(Si{\sc ii})/$N$(Al{\sc ii})] = 1.05, log[$N$(Fe{\sc ii})/$N$(Al{\sc ii})] = 1.39. For the object J113717.73$+$354956.9 log[$N$(Fe{\sc ii})/$N$(Al{\sc ii})] = 1.47. Listed are the hydrogen density ($n_\mathrm{H}$) in cm$^{-3}$, pressure ($P$) in dyne cm$^{-2}$,  Jeans length ($\lambda_\mathrm{J}$) in kpc, cloud radius ($R_c$) in kpc, cloud mass ($M_c$) in $M_\odot$, impact parameter ($b_\perp$) in kpc and $b_\text{frac}=b_\perp/R_c$ for isobaric cloud models. All models assume a UV background at $z=6$ from \citet{HM2012} and solar abundances with $Z=0.1Z_\odot$. (All column densities and their ratios are expressed as log.) 
}

    \begin{tabular}{lccccccr}
    \hline
    \hline
   SDSS QSO name & $z_\text{abs}$ & $\lambda$ & W$_0$  & $\log \lambda f$ & $\log N$(Si{\sc ii}) &  $\log N$(Fe {\sc ii}) &$\log N$(Al {\sc ii})  \\
    \hline
   J084035.09+562419.9&  5.5940& 1526.70698  & 0.52  & 2.546   & 14.039  &  $\dotsc$   &    $\dotsc$  \\
   & 5.5938 & 1608.45085  & 0.34 &  1.998 &  $\dotsc$   & 14.38  &    $\dotsc$  \\
      &5.5943 & 1670.7886  & 0.44 & 3.486  &  $\dotsc$   &  $\dotsc$   &   12.98 \\
    \hline
    J113717.73+354956.9 &5.0120 & 1608.45085  & 0.65 &1.998   &  $\dotsc$   & 14.66 &    $\dotsc$ \\
     &5.0127 & 1670.7886  & 0.7 & 3.486  &  $\dotsc$   &  $\dotsc$   & 13.18 \\
    \end{tabular}

    \begin{tabular}{lccccccccccccr}
    \hline
    \hline
  H{\sc i} & Si{\sc ii} &   Al{\sc ii} & Fe{\sc ii} &  $\mathrm{\frac{SiII}{FeII}}$ &  $\mathrm{\frac{SiII}{AlII}}$&  $\mathrm{\frac{FeII}{AlII}}$& log $n_\mathrm{H}$ &  log $P$     & $\lambda_\mathrm{J}$ &   $R_c$ &   $\log M_c$  &  $b_\perp$ &  $b_\text{frac}$ \\
\hline
18.99 & 14.30 & 13.40 & 13.93 &  0.36 & 0.89 &  0.52 &   -2.0 & -13.40 &  11.03  &   1.34 &  6.75 &  0.2    & 0.15 \\
17.76 & 13.48 & 12.57 & 12.97 &  0.50 & 0.90 &  0.40 &   -1.5 & -12.95 &   5.89  &   0.16 &  4.61 &  0.15   & 0.91 \\
19.42 & 14.21 & 13.17 & 14.03 &  0.17 & 1.03 &  0.86 &   -1.5 & -12.95 &   5.89  &   0.16 &  4.61 &  0.0    & 0.0  \\
19.21 & 13.94 & 12.87 & 13.81 &  0.12 & 1.06 &  0.94 &   -1.0 & -12.51 &   3.16  &   0.02 &  2.75 &  0.003  & 0.10 \\
19.17 & 13.80 & 12.73 & 13.70 &  0.09 & 1.06 &  0.96 &   -0.5 & -12.09 &   1.70  &   0.005 &  1.12 &  0.0    & 0.0  \\

19.54 & 14.59 & 13.83 & 14.22 &  0.37 & 0.76 &  0.38 &  -2.5 & -13.86 &   20.59 &   12.91 &  9.11 &  0.11 & 0.009  \\
19.57 & 14.60 & 13.83 & 14.23 &  0.36 & 0.76 &  0.40 &  -2.5 & -13.86 &   20.59 &   12.91 &  9.11 &  0.0 & 0.0 \\

\hline
    \end{tabular}
    \label{ModelsRyanWeber:table}
\end{table*}

For completeness, we also looked at the early data of two low-ionisation systems (showing Al{\sc ii}, Fe{\sc ii} and in one case Si{\sc ii}) detected in the spectra of QSOs at $z_\mathrm{abs}>5$ \citep{RyanWeber2009MNRAS.395.1476R}. (Six other systems at $z>5$ show only C{\sc iv}.) The measurements were taken with NIRSPEC on the Keck {\sc II} telescope and with ISAAC on the VLT-UT1, with a SNR$\gtrsim$ 5. A summary of the observations is provided in the top panel of Table~\ref{ModelsRyanWeber:table}. The reported equivalent widths were converted to column densities using the atomic data from \citet{Morton1991ApJS...77..119M}. The bottom panel shows the physical properties of representative models for the clouds. For the system at $z_\text{abs}=5.594$ in the spectrum of J084035.09+562419.9, the models struggle to reproduce the measured ratio log[$N$(Si{\sc ii})/$N$(Fe{\sc ii})] = -0.34. As uncertainties on the measurements are not provided, we cannot estimate the degree of discrepancy between the models and the measurements. The closest model we find is log[$N$(Si{\sc ii})/$N$(Fe{\sc ii})] = 0.09 (fifth model in Table~\ref{ModelsRyanWeber:table}), corresponding to a small cloud ($R_c\approx 5$~pc) with high hydrogen density and with a mass of $\sim13\, M_\odot$, well below the thermal evaporation mass limit. Models match the measured value log[$N$(Si{\sc ii})/$N$(Al{\sc ii})] $\simeq 1.06$, such as a cloud with $R_c\approx 160$~pc and mass $\sim4.1\times10^{4}\, M_\odot$. The measured value log[$N$(Fe{\sc ii})/$N$(Al {\sc ii})] = 1.4 was not matched by the models. Values above 0.94 require very small cloud sizes, below $\sim20$~pc, with high hydrogen densities and masses below the thermal evaporation limit. The H{\sc i} column densities for the models best matching the data correspond to sub-DLAs.
For the system at $z_\text{abs}=5.012$ in the spectrum of J113717.73+354956.9, the measured value log[$N$(Fe {\sc ii})/$N$(Al {\sc ii})] $\simeq 0.47$, the best match is provided by clouds with larger sizes of $\sim0.1-10$~kpc and masses $\sim10^4-10^9\, M_\odot$, and possibly with a H{\sc i} column density smaller than $10^{19}\,\text{cm}^{-2}$. It would be interesting to re-observe this system at higher resolution and signal-to-noise ratio, particularly to investigate possible sub-components of the features.

\section{Discussion}
\label{discussion}

Our main goal in this paper is to interpret the metal absorption line data in sub-DLAs and DLAs at high redshift in terms of a model of clouds pressure-confined by the gas in galactic haloes to infer the properties of the stellar populations that produced the metals. As these systems arise in the first billion years of the Universe, the metals should shed light on the stellar populations of the first galaxies. Central to this goal is establishing whether the pressure-confined model is supported by the data. In this section, we discuss our findings for the pressure-confined cloud models as they relate to individual sets of observations.

For the systems observed at these redshifts by \citet{Becker:2011, Becker2011-12} (Figs.~\ref{fig:BeckerSiIIvsCII} and \ref{fig:BeckerSiIItoCIIvsFeIItoCII}), the models are consistent with their suggestion that the low ionisation metal systems they detected arise in DLAs. Models with $\log N$(H{\sc i}$)=10^{20}-10^{21}\,\text{cm}^{-2}$ provide $N$(C{\sc ii})/ $N$(O{\sc i}) ratios consistent with their data for metal abundances between starburst and solar values and metallicities of $0.001-0.01 Z_\odot$. The measured values of $N$(Si{\sc ii})/ $N$(O{\sc i}), however, lie systematically low by about 0.2 dex for lines of site passing through the cloud centres. Boosting the UVB shifts the Si{\sc ii} to O{\sc ii} and C{\sc ii} to O{\sc ii} mostly along the same locus as varying the impact parameter, and away from the measured values. Depletion onto dust grains is another complicating factor for interpreting metal line ratios in DLAs and sub-DLAs. As silicon is a refractory element, the offset may arise from depletion within the cloud centres: for the low metallicities of these systems, very little depletion is expected, but 0.2 dex is consistent with the level of depletion assessed for low metallicity DLAs at moderate redshifts of $z\sim2.5$ \citep{2005A&A...440..499A}. Depletion analyses of $z\gtrsim5$ sub-DLAs and DLAs also support depletion levels at least as strong as at moderate redshifts for comparable metallicities \citep{2016ApJ...830..158M, Poudel2019}. Allowing for this level of silicon depletion, the measured $N$(Fe{\sc ii})/ $N$(C{\sc ii}) values again broadly bridge abundances between solar and starburst values. The measured ratio $N$(Si{\sc ii})/$N$(Fe{\sc ii}) is found enhanced over solar by about 0.25--0.45~dex. Models with enhanced UVB intensities (multiplying by factors of 2 and 4) shift the Fe{\sc ii} to Si{\sc ii} ratio by only $0.05-0.1$~dex. Since iron is also a refractory element, the larger measured shifts may be accounted for by depletion onto dust grains. As pointed out by \citet{Becker2011-12}, the shifts are also consistent with abundance measurements for low metallicity halo stars in the Galaxy \citep{2004AnA...416.1117C}. We refer to \citet{Becker2011-12} and \citet{2011MNRAS.417.1534C} for further discussion of the interpretation of metal abundances in very low metallicity DLAs.

The column density ratios for the systems measured by \citet{Poudel2018} are similar to those from the slightly higher redshift data of \citet{Becker:2011, Becker2011-12}, again consistent with abundances between starburst and solar. One system ($z_\text{abs}=5.335$) illustrates a possible hazard in interpreting under-resolved features. Voigt-profile fitting resolves the feature into two sub-components. The total metal column densities of C{\sc ii}, Si{\sc ii} and O{\sc i} are consistent with a line of sight passing through the core of a DLA with solar abundances, and inconsistent with any model with starburst abundances. The sub-components, however, tell another story. Whilst one is consistent with the solar abundance interpretation, it favours a line of sight offset from the absorber core. The C{\sc ii} to O{\sc i} ratio for the other sub-component is inconsistent with a model having solar abundances, but is consistent with a starburst abundance interpretation. Its Si{\sc ii} to O{\sc i} ratio is too low, but only by $\sim0.25$~dex, which could indicate a small level of depletion onto dust grains. The interpretation of the sub-components is thus rather different from the interpretation the averaged column densities would suggest.

Similar behaviour is found for the sub-DLA at $z_\text{abs}=4.98$ \citep{2016ApJ...830..158M}. Ratios of the total metal column densities for the feature suggest a silicon depleted system with solar abundances. The feature is resolved into several sub-components, three of which have all of O{\sc i}, C{\sc ii} and Si{\sc ii} detected. A highly silicon-depleted sub-component survives, but its abundance is less clear-cut, consistent with both solar and starburst abundances given the errors, as is another sub-component. The remaining, however, strongly favours starburst abundances, and moreover suggests the line of sight passes outside the core of the cloud. High spectral resolution data is clearly required to interpret complex metal absorption features, which may arise from a mixture of systems with varied chemical composition.

Compared with the systems above, whilst the abundances inferred from the O{\sc i}, C{\sc ii} and Si{\sc ii} column densities measured by \citet{Poudel2019} span the range from starburst to solar, they more broadly indicate starburst abundances in the clouds. Consistency with the pressure-confined models, however, suggests the lines of sight often pass outside the cloud cores, as the predicted ratios of Si{\sc ii} to O{\sc i} are otherwise too small. As shown in Fig.~\ref{fig:metIonStage}, the C{\sc ii} and Si{\sc ii} fractions fall away much more slowly than H{\sc i} and O{\sc i} for lines of sight increasingly displaced from the cloud centres. Geometry may play an important role in interpreting the metal line ratios in systems optically thick to ionising radiation. 

Almost all of the absorption systems are resolved into sub-components by Voigt-profile fitting. In every case but one, when a system contains one or more sub-components with column density ratios corresponding to a line of sight consistent with passing outside a cloud core, the complex contains at least one other sub-component favouring a line of sight passing through a cloud core. The measured H{\sc i} may be interpreted as arising from the latter, as the lines of sight passing outside the core may have $\log N$(H{\sc i})$<20$.\footnote{An alternative interpretation of column density ratios that do not correspond to the predictions for the core of a DLA or sub-DLA, where they are largely shielded from the metagalactic photoionising radiation field, is that they arise from lower H{\sc i} column density systems that need not be pressure confined. Our point is the pressure-confined models may be able to explain the column density ratios for these absorption systems as well.} An exception is the complex at $z_\text{abs}=4.987$ in QSO J0306$+$1853. \citet{Poudel2019} obtain a silicon-to-oxygen ratio for the system of [Si/O] = 0.79$\pm$0.09, which they recognise as surprisingly high. As the QSO was observed with the lowest resolution of any of their observations ($R=4100$), they suggest that higher resolution observations may be required to obtain more accurate column densities for the metals. Should higher resolution observations maintain the high silicon abundance, one interpretation may be that most of the measured H{\sc i} arises in a near pristine cloud, with a metallicity below $0.001Z_\odot$, and that the high Si{\sc ii} to O{\sc i} ratio originates from gas outside the core of a second pressure-confined cloud that has been polluted by metals (such as Model 1 in Table~\ref{ModelsPoudel:table}).

We note that a determination of the metallicities is less secure since the measured H{\sc i} column densities are strictly upper limits to the metal systems since the Lyman $\alpha$ and  Lyman $\beta$ (when available) absorption features alone are broader than the velocity separation betwen metal sub-components in a complex. Without higher order Lyman series measurements to isolate the redshift of H{\sc i} absorption corresponding to the metal features, the H{\sc i} column density to assign to each sub-component is unclear. We find models with metallicities $0.001-0.01Z_\odot$ provide good matches to the data. When Fe{\sc ii} measurements are available, the metallicities are closer to $0.01Z_\odot$. \citet{Poudel2019} infer metallicities somewhat above $0.01Z_\odot$ for some systems, but do not allow for an ionisation correction to the measured H{\sc i} column density. For a line of sight passing outside the core of a cloud, Fig. ~\ref{fig:metIonStage} shows the ionisation corrections may be substantial, as much as an order of magnitude or larger for H{\sc i} than for C{\sc ii} and Si{\sc ii}. On the other hand, the correction for oxygen is comparable to that for hydrogen (since the ionisation potentials are nearly the same), so that the ratio of O{\sc i} to H{\sc i} should be a good indicator of the metallicity, although it will be abundance-dependent if referenced to iron, which is 0.8~dex less abundant for starburst abundances than solar for the same oxygen abundance (see Table~\ref{tab:abundances}). The anomalously high iron abundance for the system at $z_\text{abs}=4.589$ of $Z_\text{Fe}\approx0.1Z_\odot$ is difficult to accommodate in any model, especially as it appears in a sub-component without other detected metal absorption based on the Voigt-profile fit to the overall feature, which was resolved into two sub-components. \citet{Poudel2019} note the iron feature is blue-shifted relative to the other metal features, which may indicate local iron-enriched material in an outflow from a Type Ia supernova. A system measured by \citet{RyanWeber2009MNRAS.395.1476R} (at $z_\text{abs}=5.594$) similarly shows an anomalously high iron abundance the models are unable to match.

We also examined the possibility that some of the high ionisation absorption systems, including C{\sc iv} and Si{\sc iv}, reported by \citet{DOdorico2013MNRAS.435.1198D} may arise in the outer layers of pressure-confined sub-DLA or DLA systems. The column density ratios do not support the possibility:\ the systems appear generally to have H{\sc i} column densities of around $10^{18}\,\text{cm}^{-2}$.

\section{Conclusions}
\label{conclusions}

The metal absorption lines measured in sub-DLA and DLA systems at $z\gtrsim5$ may provide important clues to the origin and nature of galaxies in the first billion years in the Universe. We analysed measurements reported in the literature of sub-DLA and DLA systems to test the possibility that such systems arise from pressure-confined clouds in galactic haloes, and interpret the measurements in the context of this model to infer the physical properties of the clouds. To do so, we extended simple slab models for sub-DLAs and DLAs to allow for lines of site that pass outside the largely neutral hydrogen cores of the clouds. We approximated the clouds by rolling slab models computed using {\sc cloudy} into cylinders, which should capture the main effects of ionisation layers in a spherical pressure-confined cloud on the metal column density ratios for off-centre lines of sight. Using this procedure, we found pressure-confined clouds provide a viable model for low-ionisation intervening metal absorption systems approaching the epoch of reionization.

We reached the following specific conclusions:

\begin{itemize}
\item Typical gas densities, pressures, sizes and masses of acceptable model clouds range over, respectively, $0.01\la n_\text{H}\la0.1\,\text{cm}^{-3}$, $-14.2<\log P [\text{dyne\, cm}^{-2}]<-12.5$, $0.04<R_c<3$~kpc and $3.5<\log M_c/M_\odot<8$. The cloud pressures are consistent with expectations for virialized regions of dark matter haloes at $5\la z\la6$ with masses $11<\log M_h/h^{-1}M_\odot < 12$. The gas densities must exceed $n_\text{H}>0.001\,\text{cm}^{-3}$ for the clouds to be Jeans stable.
\item The best-fitting models have typical H{\sc i} column densities consistent with sub-DLAs ($10^{19}-10^{20.3}\,\text{cm}^{-2}$) and DLAs ($>10^{20.3}\,\text{cm}^{-2}$), with metallicities $0.001-0.01Z_\odot$. 
\item The best-fitting models have metal abundances that range between $\alpha$-enhanced abundances expected for a stellar population dominated by massive stars \citep[startburst abundances based on a chemical evolution model from][with a top-heavy stellar initial mass function]{HamannFerland:1993} and solar abundances. 
\item The best-fitting model inferred from the ratios between total metal column densities in an absorption complex can differ qualitatively from the models that best fit the individual sub-components, emphasising the need for high resolution, high signal-to-noise ratio data.
\item The model predicts that any sub-component arising from a line of sight offset from the core of a cloud should be accompanied by at least one sub-component arising from a line of sight passing through a cloud core to account for the high H{\sc i} column density. This is found generally borne out by the data.
\item Models of the two proximate DLAs \citep{D_Odorico_2018, Banados2019ApJ...885...59B} examined favour solar abundances, although the systems are too few to draw a general conclusion.

\item Variations in the UVB have a moderate effect on the metal ion column densities, but primarily for lines-of-sight in the outer regions of the clouds where the absorption of H{\sc i} or O{\sc i} ionising radiation is reduced. The shifts in the column densities of singly ionised species is only about $0.05-0.1$~dex.
\end{itemize}

The masses and radii estimated for the pressure-confined systems reproducing the metal absorption data make it challenging to study the clouds using cosmological hydro-simulations. Recent simulations exploring the impact spatial and mass resolution has on the modelling of gas clouds and filaments in galaxy scale simulations are just beginning to reach the scales of the moderately sized clouds, but resolving scales down to 10~pc is still beyond their capacity \citep[eg][]{Peeples2019, 2019MNRAS.482L..85V}. Resolving the sizes of low-ionisation metal species is especially difficult. Establishing the relation of low-ionisation clouds to galaxy formation must continue to await further improvements in the numerical simulations. Until then, analytic models provide a helpful means of interpreting the growing amount of data.

\section*{Acknowledgements}
TS and AM gratefully acknowledge support from the UK Science and Technology Facilities Council, Consolidate Grant ST/R000972/1. TS thanks Juan V. Hern\'andez for useful coding comments. 
This research also made use of {\sc astropy}, a community-developed core {\sc python} package for Astronomy \citep{Astropy-Collaboration:2013aa} and {\sc matplotlib} \citep{Hunter:2007aa}.

\section*{Data Availability}
The data underlying this article will be shared on reasonable request to the corresponding author.



\bibliographystyle{mnras}
\bibliography{bibliography} 




\appendix

\section{Tables}
\label{obs:tables}
In this section, tables are provided of the data used for the figures taken from \citet{Becker:2011}, \citet{Becker2011-12}, \citet{2016ApJ...830..158M}, \citet{Poudel2018} and \citet{Poudel2019}, \citet{DOdorico2013MNRAS.435.1198D} and \citet{Banados2019ApJ...885...59B}.

\begin{table*}
    \centering
    \caption{Data from \citet{Becker:2011, Becker2011-12}}
    \label{Becker:Table3}
    \begin{tabular}{lccccccr}
    \hline
    \hline
 SDSS QSO name  & $z_\mathrm{abs}$ & $\log N_\mathrm{O I}$ & $\log N_\mathrm{C II}$  & $\log N_\mathrm{SiII}$ & $\log N_\mathrm{Fe II}$ & $\log N_\mathrm{C IV}$  & $\log N_\mathrm{Si IV}$

\cr
\hline

J0040$-$0915 & 4.7393 &	$>$ 15.0     &	$>$14.6      &	14.13$\pm$0.02 &	13.77$\pm$0.06	&	  $\dotsc$  &	$\dotsc$
\cr
J1208+0010 & 5.0817 &	$>$ 14.7     &	$>$14.3      &	13.75$\pm$0.03 &	13.27$\pm$0.07	&	  $\dotsc$   &	$\dotsc$
\cr	
J0231$-$0728 & 5.338	&	14.47$\pm$0.05 &	13.79$\pm$0.05 &	13.15$\pm$0.04 &	12.73$\pm$0.04	&	  $\dotsc$  &	$\dotsc$
\cr
       	   &	    & $\dotsc$ 	 &	14.13$\pm$0.04 &	13.42$\pm$0.02 &12.94$\pm$0.05 &		   	  $\dotsc$   &	$\dotsc$
\cr
J0818+1722 & 5.7911 &	14.54$\pm$0.03 &	14.13$\pm$0.03 &	13.36$\pm$0.04 &	12.89$\pm$0.07 & $\dotsc$  & $\dotsc$ 
\cr
	       &	    & $\dotsc$ 	 &	14.19$\pm$0.03 &	13.41$\pm$0.03 &	12.95$\pm$0.07		    & 	$<$13.4	  & $<$12.9
\cr
J0818+1722 & 5.8765 &	14.04$\pm$0.05 &	13.62$\pm$0.07 &	12.78$\pm$0.05 &	<12.7	    &	$<$13.0 &  $<$12.6
\cr
J1623+3112 & 5.8415 &	$>$15.0 &	$>$14.3 &	14.09$\pm$0.02 &	$\dotsc$	    &	$\dotsc$ &  $<$13.0 
\cr
J1148+5251 & 6.0115 &	14.65$\pm$0.02 &	14.14$\pm$0.06 &	13.51$\pm$0.03 &	13.21$\pm$0.25	&	$<$13.8 & $<$12.7
\cr
J1148+5251 & 6.1312 &	14.79$\pm$0.23 &	13.88$\pm$0.15 &	13.29$\pm$0.06 &	13.03$\pm$0.18	&	$<$13.2 & $<$12.3
\cr
J1148+5251 & 6.1988 &	13.49$\pm$0.13 &	12.90$\pm$0.11 &	12.12$\pm$0.05 &	 $\dotsc$ &	$<$13.6 & $<$12.9
\cr
J1148+5251 & 6.2575 &	14.12$\pm$0.15 &	13.78$\pm$0.21 &	12.92$\pm$0.08 &	$<$12.7	    &	$\dotsc$     & $<$12.5
\cr
\hline
    \end{tabular}
\end{table*}

\begin{table*}
    \centering 
    \caption{Data from \citet{2016ApJ...830..158M}.}
    \label{Morrison:table01}
    \begin{tabular}{lccccccr}
    \hline 
    \hline 
 SDSS QSO name  & $z_\mathrm{abs}$& $z$ &  $\log N_\mathrm{H I}$&  $\log N_\mathrm{O I}$ & $\log N_\mathrm{C II}$  & $\log N_\mathrm{SiII}$  & $\log N_\mathrm{FeII}$  

\cr 
\hline 
Q1202$+$3235	& z = 4.977	&	4.977004$\pm$0.000002	& $\dotsc$ &		14.50$\pm$0.08 &	13.91$\pm$0.08 & 12.89$\pm$0.04  & 12.86$\pm$0.12
\cr 
	&	&	4.977259$\pm$0.000009	&$\dotsc$ &	13.07$\pm$0.21	& 12.63$\pm$ 0.28 &	12.39$\pm$0.13 & $\dotsc$
\cr	
	&	&	4.978517$\pm$0.000008	&$\dotsc$ &	12.89$\pm$0.10	& 13.66$\pm$ 0.07 &	12.79$\pm$0.05 & $\dotsc$
\cr	
	&	&	4.978761$\pm$0.000008	&$\dotsc$ &	13.02$\pm$0.07	& 14.00$\pm$ 0.12 &	$\dotsc$ & $\dotsc$
\cr	
	& & Total log N	& 19.83$\pm$0.10 &	14.54$\pm$0.07	&	14.48$\pm$0.05	& 13.21$\pm$0.03 & $12.86\pm$0.12
\cr 
\hline 
\hline 
    \end{tabular}
\end{table*}

\begin{table*}
    \centering 
    \caption{Data from \citet{Poudel2018}.}
    \label{Poudel:table01}
    \begin{tabular}{lccccccr}
    \hline 
    \hline 
 SDSS QSO name  & $z_\mathrm{abs}$& $z$ &  $\log N_\mathrm{H I}$&  $\log N_\mathrm{O I}$ & $\log N_\mathrm{C II}$  & $\log N_\mathrm{SiII}$  & $\log N_\mathrm{FeII}$  

\cr 
\hline 
Q0231$-$0728	& z = 5.335	&	5.33505$\pm$0.00003	& $\dotsc$ &		14.41$\pm$0.05	& 13.54$\pm$0.10  &	12.91$\pm$0.06 & $\dotsc$
\cr 
	&	&	5.33636$\pm$0.00015	&$\dotsc$ &	13.89$\pm$0.15	& 13.83$\pm$ 0.07 &	13.08$\pm$0.05 & $\dotsc$
\cr	
	& & Total log N	& 20.10$\pm$0.15 &	14.55$\pm$0.05	&	14.18$\pm$0.05	& 13.39$\pm$0.03 & $\dotsc$
\cr 
\hline
Q0824+1302	& z = 4.809	&	4.80253$\pm$0.00013	&$\dotsc$ &	13.01$\pm$0.29	& 13.13$\pm$0.19	& 12.74$\pm$0.28 & $\dotsc$
\cr
	&	&	4.80922$\pm$0.00003	& $\dotsc$ &	14.18$\pm$0.06	& 13.62$\pm$0.10	& 13.16$\pm$0.16 & $\dotsc$
\cr
	& & Total log N	& 20.10$\pm$0.15&	14.28$\pm$0.06	&	13.77$\pm$0.09	& 13.30$\pm$0.14 & 13.12$\pm$0.17
\cr
\hline

Q0824+1302  & z = 4.829 &	4.82908$\pm$0.00002	&$\dotsc$ &	15.44$\pm$0.15	& 14.88$\pm$0.12	& 14.17$\pm$0.09 & $\dotsc$
\cr
	& & Total log N	& 20.80$\pm$0.15 &	15.44$\pm$0.15	&	14.88$\pm$0.12	& 14.22$\pm$0.08 & 13.82$\pm$0.13
\cr
\hline 
\hline 
    \end{tabular}
\end{table*}

\begin{table*}
    \centering
    \caption{Data from Table 2 of \citet{Poudel2019}. }
    \label{Poudel:tables02}
    \begin{tabular}{lccccccr}
    \hline
    \hline

SDSS QSO name  & $z_\mathrm{abs}$& $z$ & $\log N_\mathrm{HI}$ & $\log N_\mathrm{OI}$ & $\log N_\mathrm{CII}$  & $\log N_\mathrm{SiII}$ & $\log N_\mathrm{FeII}$
\cr
\hline

J0306+1853	&	4.987 	& 	4.98661$\pm$0.00004	&	$\dotsc$	& 14.28$\pm$0.16	&	14.13$\pm$0.15	&	13.9$\pm$0.07	& $\dotsc$	 \cr
&	& 	Total log N	&	20.60$\pm$0.15	&	14.60$\pm$0.08 & 14.37$\pm$0.13 & 14.22$\pm$0.05	& $\dotsc$ \cr
\hline 
 J1233+0622	&	4.859 	& 	4.85795$\pm$0.00004	&	$\dotsc$	& $>15.30$	&	14.42$\pm$0.07	&	13.82$\pm$0.19	&	$\dotsc$ \cr 
&	& 	Total log N	&	20.75$\pm$0.15	&	$>$15.30 & 14.74$\pm$0.05 & 14.35 $\pm$ 0.10	& 13.61$\pm$0.13  \cr		
J1233+0622*	&	5.050	& 	5.05024$\pm$0.00001	&	$\dotsc$	& 15.11$\pm$0.12	&	14.75$\pm$0.16	&	14.19$\pm$0.11	& 14.07$\pm$0.09 \cr
 	&		& 	5.05188$\pm$0.00004	&	$\dotsc$	& 13.88$\pm$0.19	& 13.84$\pm$0.19	&	13.69$\pm$0.17	&	  $\dotsc$ \cr
	&	& 	Total log N	&	20.10$\pm$0.15  & 15.19 $\pm$ 0.10 & 14.93 $\pm$ 0.11 & 14.31 $\pm$ 0.09 & 14.07 $\pm$ 0.09\cr
\hline
J1253+1046	&	4.589 	& 	4.58916$\pm$0.00005	&	$\dotsc$	&	$\dotsc$	&	$\dotsc$	&	$\dotsc$	&	13.89$\pm$0.17 \cr
&		& 	4.58948$\pm$0.00004	&	$\dotsc$	&	14.94$\pm$0.10	&	14.34$\pm$0.22	&	13.67$\pm$0.09	&	$\dotsc$ \cr
	&	& 	Total log N	&	19.75$\pm$0.15 & 15.00 $\pm$ 0.08 & 14.34 $\pm$ 0.22 & 13.67 $\pm$ 0.09 & 14.24 $\pm$ 0.11 \cr			
	&	4.600 	& 	4.60003$\pm$0.00003	&	$\dotsc$	&	15.53$\pm$0.08	&	14.78$\pm$0.20	&	14.57$\pm$0.05	&   $\dotsc$ \cr	
	&	& 	Total log N	& 20.35$\pm$0.15 & $>$ 15.58 & $>$ 14.78 & 14.57$\pm$0.05 & 14.06 $\pm$ 0.29 \cr
\hline
J1557+1018	&	4.627 	& 	4.62512$\pm$0.00004	&	$\dotsc$	&	14.54$\pm$0.11	&	14.38$\pm$0.15	&	13.65$\pm$0.06	&	  $\dotsc$ \cr 
	&		& 	4.62694$\pm$0.00002	&	$\dotsc$	&	15.64$\pm$0.09	&	15.52$\pm$0.11	&	15.00$\pm$0.04	&	 $\dotsc$ \cr 
	&	 	& 	4.62885$\pm$0.00003	&	$\dotsc$	&	15.55$\pm$0.19	&	15.10$\pm$0.21	&	14.74$\pm$0.04	&	  $\dotsc$ \cr 
	&	 	& 	4.63018$\pm$0.00015	&	$\dotsc$	&	15.01$\pm$0.19	&	14.53$\pm$0.19	&	14.35$\pm$0.09	&	 $\dotsc$ \cr 
	&	& 	Total log N	&20.75$\pm$0.15&  $>$ 15.97 & $>$ 15.71 & $>$ 15.26 & $\dotsc$ \cr

\hline
    \end{tabular}
\end{table*}

\begin{table*}
    \centering
    \caption{Data from \citet{DOdorico2013MNRAS.435.1198D}.}
    \label{Dodorico}
    \begin{tabular}{lcccc}
    \hline
    \hline
 SDSS QSO name  & $z_\mathrm{abs}$ & $\log N_\mathrm{C II}$ &  $\log N_\mathrm{C IV}$  & $\log N_\mathrm{Si IV}$
\cr
\hline
SDSSJ0818$+$1722    &   5.7899$\pm$0.0002   &    $\dotsc$               &   13.2$\pm$0.1    &   $<12.5$     \\
                    &   5.78909$\pm$0.00004 &   13.56$\pm$0.04  &     $\dotsc$            &   $\dotsc$  \\
\cr
                    &   5.8441$\pm$0.0001   &   $>$ 13.0        &   13.27$\pm$0.07  &   12.69$\pm$0.06  \\
\cr
                    &   5.8770$\pm$0.0001   &     $\dotsc$            &   13.22$\pm$0.07  &   $<$12.5\\
                    &   5.87644$\pm$00002   &   13.81$\pm$0.03  &        $\dotsc$     &      $\dotsc$    \\      
\hline
SDSSJ0836$+$0054    &   5.32277$\pm$0.00004 &   $<$13.0         &   13.65$\pm$0.04  &   12.77$\pm$0.04\\
\hline
SDSSJ1030$+$0524    &   5.72419$\pm$0.0001  &   $<$13.5         &   14.52$\pm$0.08  &   $<$13.7 \\
\cr
                    &   5.74116$\pm$0.00004 &         $\dotsc$     &   13.8$\pm$0.1    &   13.20$\pm$0.03\\
                    &   5.74097$\pm$0.00001 &   14.40$\pm$0.01  &          $\dotsc$     &        $\dotsc$     \\
\cr
                    &   5.7440$\pm$0.0002   &          $\dotsc$     &   13.89$\pm$0.09  &          $\dotsc$    \\
                    &   5.74425$\pm$0.00004 &         $\dotsc$    &       $\dotsc$    &   13.34$\pm$0.03  \\          
                    &   5.74399$\pm$0.00004 &   14.34$\pm$0.01  &      $\dotsc$      &      $\dotsc$     \\
\cr
                    &   5.9757$\pm$0.0004   &   $<$13.5         &   13.1$\pm$0.3    &   $<$12.7         \\
\cr
                    &   5.9784$\pm$0.0002   &        $\dotsc$    &   13.4$\pm$0.2    &        $\dotsc$    \\
                    &   5.97896$\pm$0.00009 &   $<$13.5         &        $\dotsc$     &   12.6$\pm$0.1    \\
\hline
SDSSJI319$+$0950    &   5.57049$\pm$0.00003 &   $<$13.5         &   13.97$\pm$0.10  &   13.04$\pm$0.02 \\
\cr
                    &   5.5740$\pm$0.0001   &       $\dotsc$       &   14.09$\pm$0.08  &         $\dotsc$    \\
                    &   5.57358$\pm$0.00001 &         $\dotsc$     &         $\dotsc$     &   13.52$\pm$0.01  \\
                    &   5.57372$\pm$0.00001 &   13.88$\pm$0.02  &        $\dotsc$     &       $\dotsc$     \\
\cr
CFHQS J1509$-$1749  &   5.91572$\pm$0.00006 &   $<$13.4         &   14.11$\pm$0.14  &   13.04$\pm$0.06  \\
\hline
    \end{tabular}
\end{table*}

\begin{table*}
    \centering
    \caption{\textit{Top panel}: Measurements identified in the DLA at $z = 6.40392$ towards QSO P183$+$ 05 at $z=6.4386$ reported in \citet{Banados2019ApJ...885...59B}. The value of Si{\sc ii} is considered to be an upper-limit, for which we have taken the highest value reported in Table 1 of \citet{Banados2019ApJ...885...59B}. The errors in parentheses are formal; the authors provide the larger more conservative errors to allow for the possibility the lines are saturated.
    \textit{Middle panel}: Estimated column densities and ratios from ionisation models in {\sc cloudy}. (The metal column densities are normalised to $0.001Z_{\odot, Si}$ for both solar and starburst abundance models.)
    \textit{Bottom panel}: Physical properties and parameters corresponding to the systems in the middle panel. The \citet{HM2012} UV background at $z=6$ has been multiplied by the indicated UVB factor.
    }
       \begin{tabular}{cccccc}
    \hline
    \hline
  $\log N$(C II) &   $\log N$(O{\sc i)}  &  $\log N$(Si{\sc ii}) &  $\log N$(Fe{\sc ii}) & $\log N$(Al{\sc ii}) & $\log N$(Mg{\sc ii}) \\ 
    \hline
14.30$\pm$0.05 &	14.45$\pm$0.20 (0.06)	& 14.15$\pm$0.05	& 13.19$\pm$0.05 & 12.49$\pm$0.20 (0.06)&  13.37$\pm$0.04 \\

    \end{tabular}

\begin{tabular}{rccccccccccccc}
\hline
\hline
Model  & H{\sc i} &       C{\sc ii} &        O{\sc i} &    Si{\sc ii }&   Al{\sc ii} &  Fe{\sc ii} &  Mg{\sc ii}   &  C{\sc ii}/O{\sc i} &  Si{\sc ii}/O{\sc i} &  Fe{\sc ii}/C{\sc ii} &  Si{\sc ii}/C{\sc ii} &  Fe{\sc ii}/Al{\sc ii} &  Fe{\sc ii}/Mg{\sc ii} \\
\hline

1&  19.03 & 12.80 & 12.80 & 12.59 &   $\dots$ & 11.32 & 12.34 &   -0.00 &    -0.21 &    -1.48 &     -0.215 &     $\dots$ &       -1.02 \\
2&  20.12 & 13.20 & 13.94 & 12.88 &   $\dots$ & 11.97 & 12.79 &   -0.74 &    -1.05 &    -1.22 &     -0.311 &     $\dots$ &       -0.81 \\
3&  20.30 & 13.31 & 14.12 & 12.99 &   $\dots$ & 12.11 & 12.92 &   -0.80 &    -1.13 &    -1.19 &     -0.327 &     $\dots$ &       -0.80 \\

4& 18.85 & 13.21 & 12.53 & 12.46 & 11.73 & 11.89 & 12.28 &  0.68 & -0.07 &  -1.32 &  -0.75 &    0.16 &   -0.39 \\
5& 20.30 & 13.78 & 13.99 & 12.95 & 11.99 & 12.79 & 12.91 & -0.20 & -1.03 &  -0.99 &  -0.83 &    0.79 &   -0.12 \\

6&  21.30 & 14.22 & 15.11 & 13.88 &  $\dots$  & 13.07 & 13.83 & -0.89 & -1.23 &  -1.14 &  -0.33 &   $\dots$  & -0.75 \\
7&  20.66 & 13.70 & 14.47 & 13.45 &  $\dots$  & 12.51 & 13.25 & -0.77 & -1.02 &  -1.18 &  -0.24 &   $\dots$  & -0.74 \\

\end{tabular}

\begin{tabular}{rcccccccccc}
\hline
\hline
Model & $\lambda_\mathrm{J}$ &  $n_\mathrm{H}$ &  $\log P$ &  $R_c $ &   $M_c$ &  $b_\perp$ &  $b_\mathrm{frac}$ & UVB factor & abundance & $\log N$(H{\sc i}) \\
\hline
1&     12.42  &  0.01   &   -13.30 &      6.01 &      8.83 &   0.94 &  0.16   &  $\times$4     &   Sb      &       20 \\
2&     12.42  &  0.01   &   -13.30 &      6.01 &      8.83 &   0.47 &  0.08   &  $\times$4     &   Sb      &       20 \\
3&     12.42  &  0.01   &   -13.30 &      6.01 &      8.83 &   0.00 &  0.00   &  $\times$4     &   Sb      &       20 \\

4&     11.91  &  0.01   &   -13.33 &      3.23 &      8.13 &    0.97 &    0.30 &  $\times$2     &  $\odot$  &       20 \\
5&     11.91  &  0.01   &   -13.33 &      3.23 &      8.13 &    0.03 &    0.01 &  $\times$2     &  $\odot$  &       20 \\

6&     46.36  &  0.001   &   -14.16 &    721.97 &     14.06 &    2.91 &    0.004 &  $\times$4     &   Sb      &       21 \\
7&     46.36  &  0.001   &   -14.16 &    721.97 &     14.06 &   50.55 &    0.07  &  $\times$4     &   Sb      &       21 \\

\hline
\end{tabular}

    \label{tab:banados2020}
\end{table*}


\bsp	
\label{lastpage}
\end{document}